\documentclass[twocolumn,amssymb, nobibnotes, aps, superscriptaddress]{revtex4-1}
\usepackage{amsmath}
\usepackage{graphicx}
\usepackage{bm}
\usepackage{graphicx}
\usepackage{float} 
\usepackage{subfigure}
\usepackage{makecell}
\usepackage{booktabs}
\usepackage{natbib}
\usepackage{color}
\usepackage[colorlinks=true, allcolors=blue]{hyperref}
\usepackage{wasysym}
\usepackage[utf8]{inputenc}

\begin{document}
\title{Confined Electrons in Effective Plane Fractals}%
\author{Xiaotian Yang}
\affiliation{Key Laboratory of Artificial Micro- and Nano-structures of Ministry of Education and School of Physics and Technology, Wuhan University, Wuhan 430072, China}
\author{Weiqing Zhou}
\affiliation{Key Laboratory of Artificial Micro- and Nano-structures of Ministry of Education and School of Physics and Technology, Wuhan University, Wuhan 430072, China}
\author{Peiliang Zhao}
\affiliation{Key Laboratory of Artificial Micro- and Nano-structures of Ministry of Education and School of Physics and Technology, Wuhan University, Wuhan 430072, China}
\author{Shengjun Yuan}%
\email{s.yuan@whu.edu.cn}
\affiliation{Key Laboratory of Artificial Micro- and Nano-structures of Ministry of Education and School of Physics and Technology, Wuhan University, Wuhan 430072, China}
\date{\today}
\begin{abstract}
As an emerging complex two-dimensional structure, plane fractal has attracted much attention due to its novel dimension-related physical properties. In this paper, we check the feasibility to create an effective Sierpinski carpet (SC), a plane fractal with Hausdorff dimension intermediate between one and two, by applying an external electric field to a square or a honeycomb lattice. The electric field forms a fractal geometry but the atomic structure of the underlying lattice remains the same. By calculating and comparing various electronic properties, we find parts of the electrons can be confined effectively in a fractional dimension with a relatively small field, and representing properties are very close to these in a real fractal. In particular, compared to the square lattice, the external field required to effectively confine the electron is smaller in the honeycomb lattice, suggesting that a graphene-like system will be an ideal platform to construct an effective SC experimentally. Our work paves a new way to build fractals from a top-down perspective and can motivate more studies of fractional dimensions in real systems.
\end{abstract}
\maketitle

\section{INTRODUCTION}
Different from crystal with translation symmetry, a fractal has a self-similar hierarchical structure, where the whole system is exactly or approximately similar to a part of itself.
The most prominent mathematical feature of fractals is the non-integer Hausdorff dimension $d_{H}$ \cite{geometryoffractal,feder2013fractals}. Recent experimental developments in nanofabrication, including artificial lattice fabrications \cite{gomes2012designer,polini2013artificial, slot2017experimental}, nanolithography \cite{scarabelli2015fabrication}, and etching methods \cite{de2010delocalized,nadvornik2012laterally} provide an opportunity to create high-quality arbitrary non-periodic two-dimensional (2D) structures, like plane fractals.
Practically, nanometer-scale Sierpinski hexagonal gasket \cite{newkome2006nanoassembly} and Sierpinski triangle gasket \cite{shang2015assembling,zhang2016robust,kempkes2019design} have been achieved by molecular self-assembly, chemical reaction method, and atomic manipulation.
These advances have also promoted theoretical research in this field, showing fascinating electronic and optical properties. For examples, the box-counting dimension of the quantum conductance fluctuations in the Sierpinski carpet (SC) is proved to be relevant to the Hausdorff dimension of its geometry \cite{transport}, and the sharp peaks appear in the optical spectrum due to electronic transitions between a set of specific state pairs confined in the SC at specific length scales \cite{opticalconductivity}.
More theoretical calculations about electronic transport \cite{han2019universal,bouzerar2020quantum}, quantum Hall effect \cite{iliasov2020hall,fremling2020existence}, plasmon \cite{westerhout2018plasmon}, flat bands \cite{nandy2015engineering,nandy2015flat,pal2018flat}, energy spectrum statistics \cite{iliasov2019power}, and topological properties \cite{brzezinska2018topology,pai2019topological,manna2020anyons} suggesting its future potential applications in electronics and optoelectronics.

Currently, fractals are mainly constructed by bottom-up nanofabrication methods where the system is assembled with atoms and molecules as a building unit \cite{newkome2006nanoassembly,shang2015assembling,zhang2016robust,kempkes2019design}.
However, these bottom-up approaches are limited by the number of iterations and not appreciated to fabricate large fractals.
Different from bottom-up direct synthesis, plane fractals may be built from crystal by external modification. For example, antidot lattices, which have periodic arrays of holes in graphene and other two-dimensional materials \cite{pedersen2008graphene,pedersen2008optical,furst2009electronic,gunst2011thermoelectric,yuan2013electronic,mandal2013effects,cupo2017periodic}, can be created by employing electron beam evaporation, electron beam lithography and ion milling tools \cite{eroms2009weak,mackenzie2017graphene,xu2019detecting,choudhury2019controlled}. It is a natural extension of this methodology to construct plane fractals from two-dimensional materials. However, the main limitation of this approach is that the size and position of holes are difficult to be controlled precisely. On the other hands, instead of removing sites/atoms to create holes, if there is an energy barrier between the hole and non-hole regions, electrons belong to the original hole region can not hop to the non-hole region, so that electrons in the non-hole region will be confined effectively in a space with fractal geometry. This top-down approach might be possible to form an effective fractal from two-dimensional materials with surface doping \cite{terrones2012role,wang2012doping,komsa2012two,feng2017doping,zhao2013local} or external electric fields \cite{zhao2011electronic,padilha2015van,santos2013electric,chen2020electrostatic,huber2020tunable}.
In this paper, we will check numerically the possibility to create an effective fractional space for electrons without destroying the atomic structure of the underlying crystal. To this end, we calculate the electronic properties of electrons in effective fractals with different external electric fields, and compare them to the corresponding real fractals, allowing us to figure out the minimum requirement to form a fractional space for electrons. 
Our theoretical results will provide useful information for the construction of large-scale fractals via a top-down approach.

The paper is organized as follows. In Sec.\ref{sec:model}, we describe the theoretical model and details of our numerical methods. In Sec.\ref{sec:result}, we perform the calculations of various electronic properties of different effective fractals, including Density of states (DOS), quasi-eigenstates, quantum conductance, and the box-counting dimension of the conductance fluctuation, and compare these properties with real fractals. A brief summary is given in Sec.\ref{sec:conclusion}.

\begin{figure}[H]
\centering
\includegraphics[width=15.4cm]{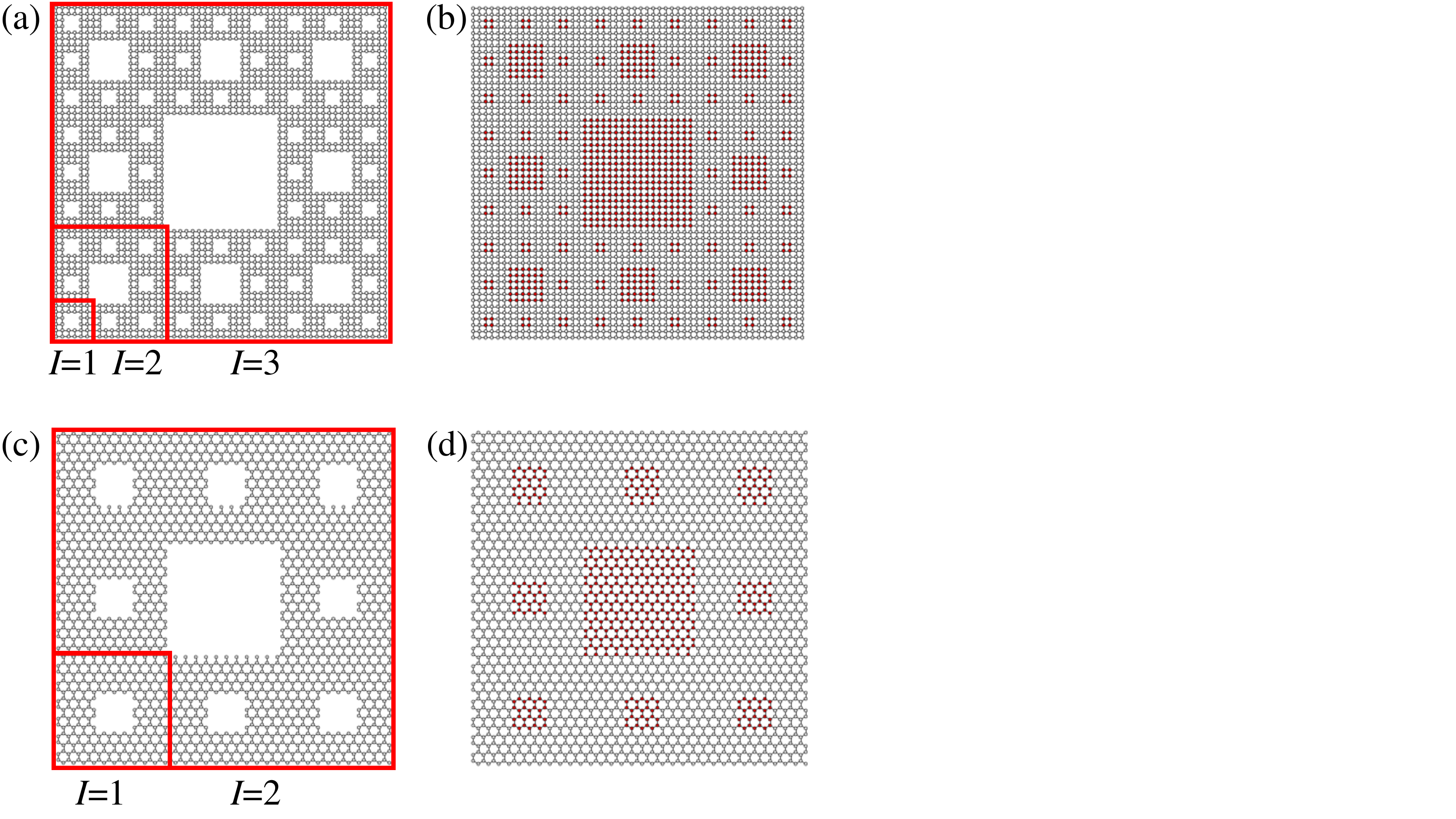}
    \caption{Schematic representation of the square-lattice Sierpinski carpet (a) and the honeycomb-lattice Sierpinski carpet (c) with different iteration $I$. The width of the sample is $W$ lattice cells (e.g. (a) $I=3$, $W=54$; (c)$I=2$, $W=33$). (b,d) illustrate the corresponding effective SC where we add opposite on-site potential in the different regions, namely adding $-V$ in the uncolored region and $V$ in the colored (red) one. Here we label the uncolored region as Area I, and the colored region as Area II in context.}
\label{sample}
\end{figure}

\section{MODEL AND METHODS}\label{sec:model}
We model a system described by a single-orbital tight-binding Hamiltonian of the form:
\begin{equation}
\begin{aligned}
H = -\sum_{\mathbf{<i,j>}}t_{ij}c^{\dag}_{i}c_{j}+\sum_{i}V_{i}c^{\dag}_{i}c_{i}.
\end{aligned}
\label{equation1}
\end{equation}
where $t_{ij}$ is electron hopping between two nearest-neighbor $i$-th and $j$-th sites and $V_{i}$ denotes on-site potential of $i$-th site. $c_i^{\dag}$ and $c_j$ are creation and annihilation operators. The geometry of square- and honeycomb-lattice based Sierpinski carpets are shown in Fig.~\ref{sample}(a) and (c), respectively. Here we introduce notations SC-$\square$ and SC-$\varhexagon$ to denote the square- and honeycomb-lattice based Sierpinski carpets. When SC changes from $I$th to $(I+1)$th iteration, the unit is replicated ${\cal N} = 8$ times larger in area and ${\cal L} = 3$ times larger in width. The Hausdorff dimension is $d_{\rm H} = \log_{{\cal L}}{\cal N}$ $\simeq 1.89$. The effective SC-$\square$ and SC-$\varhexagon$ are obtained by introducing position-dependent on-site potentials in a rectangular square or honeycomb lattice with the same width ($W$) as their corresponding real SCs. The potentials are introduced in the following way: if the site belongs to the hole region in real SC, we set $V_{i}=V$; otherwise, $V_{i}=-V$. In this way, a structure with a fractal geometry is formed in the region where all on-site potentials are $-V$ (see the uncolored Area I in Fig.~\ref{sample}(b) and (d)), which is distinguished from the hole region (Area II) where the on-site potentials are $V$.
In the limit $V\rightarrow\infty$, we expect that the states in the two regions with $V_{i}=-V$ and $V_{i}=V$ will be separated completely in energy spectrum. However, it is not clear whether it is possible to confine electrons effectively in a fractional dimension if $V$ is finite. It is therefore important to calculate the properties of the effective fractals with various finite $V$ and compare them to real fractals. We want to emphasis that, although here we add $V_{i}=-V$ in Area I and $V_{i}=V$ in Area II, one can also switch the values of potentials between two areas, namely, $V_{i}=V$ in Area I and $V_{i}=-V$ in Area II, then there will be an exchange of the energy spectrum respect to $E=0$. Furthermore, one can also choose to add only $V_{i}=-2V$ in Area I and keep the on-site potential in Area II unchanged as zero, or only add $V_{i}=2V$ in Area II. These different settings of on-site potentials only shift the whole spectrum with a constant value in the energy spectrum, leaving no change of the electronic properties. To realize these structures in the experiments, it should be easier to manipulate just one area of the sample, either Area I or II. Practically, we suggest applying an electric field in Area II (the original hole region), as it contains much less sites comparing to Area I (the region with fractal geometry).

In order to check the validity of proposed effective fractals, we will calculate their electronic and transport properties with different but finite $V$. As a comparison, the properties of corresponding real fractals with the same size would be calculated as well. As some of the systems considered in this paper contain a very large number of sites, it is numerically expensive to do all calculations based on diagonalization. Thus, we use the so-called tight-binding propagation method (TBPM) to calculate the electronic properties, including the density of states (DOS) and quasi-eigenstates, which are superpositions degenerated energy eigenstates \cite{TBPM}.
TBPM allows us to carry out calculations for rather large systems, up to hundreds of millions of sites, with a computational effort that increases only linearly with system size. The DOS is calculated via the Fourier transform of the correlation function:
\begin{equation}
\begin{aligned}
D(E)=\frac{1}{2\pi}\int_{-\infty}^{\infty}e^{iEt}\langle\varphi(0)|e^{-iHt}|\varphi(0)\rangle dt.
\end{aligned}
\label{equation2}
\end{equation}
where $|\varphi(0)\rangle$ is an initial state defined by normalized random superposition of all basis states $\sum_{n}A_{n}|n\rangle$ \cite{hams2000fast}.
The quasi-eigenstates are obtained by using the spectrum method \cite{kosloff1983fourier}. 
After the Fourier transform of states at a different time during the evolution $|\varphi(t)\rangle = e^{-iHt}|\varphi(0)\rangle$, it can be represented by \cite{TBPM}
\begin{equation}
\begin{aligned}
|\psi(E)\rangle=\frac{1}{\sqrt{\sum_{n}|A_{n}|^{2}\delta(E-E_{n})}}\sum_{n}A_{n}\delta(E-E_{n})|n\rangle.
\end{aligned}
\label{equation3}
\end{equation}
For a finite system as fractal, one can make an average from different realizations of random coefficients $A_{n}$ to get more accurate results of DOS and quasi-eigenstates.

For the transport properties, we use a quantum transport simulator KWANT to do the numerical calculations \cite{groth2014kwant}. In KWANT, the system considered is treated as a scattering region. The conductance is obtained from the scattering matrix $S_{ij}$, using the Landauer formula:
\begin{equation}
\begin{aligned}
G_{ab}=\frac{e^2}{h}\sum_{i\in{a},j\in{b}}|S_{ij}|^2 .
\end{aligned}
\label{equation4}
\end{equation}
where $a$ and $b$ refer two electrodes.

\begin{figure*}[tbp]
\centering
\includegraphics[width=16cm]{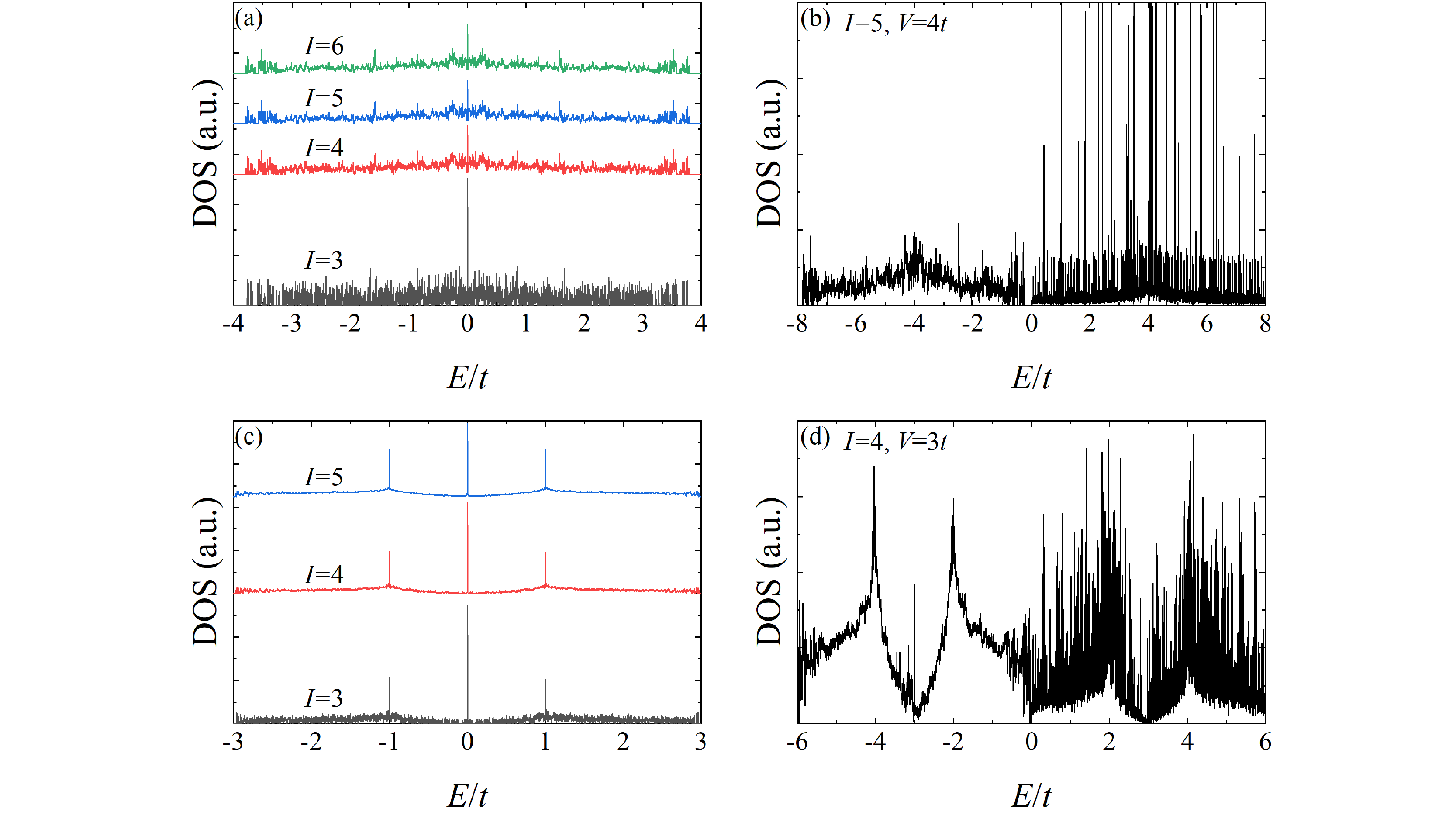}
\caption{The DOS of SC-$\square$ (a) and SC-$\varhexagon$ (c) in different iterations. (b) The DOS of the effective SC-$\square$ for $I=5$, $W=486$ and on-site potential $V=4t$. (d) The DOS of the effective SC-$\varhexagon$ for $I=4$, $W=298$ and on-site potential $V=3t$.}
\label{DOS_converged}
\end{figure*}

\begin{figure*}[tbp]
\centering
\includegraphics[width=0.8\textwidth]{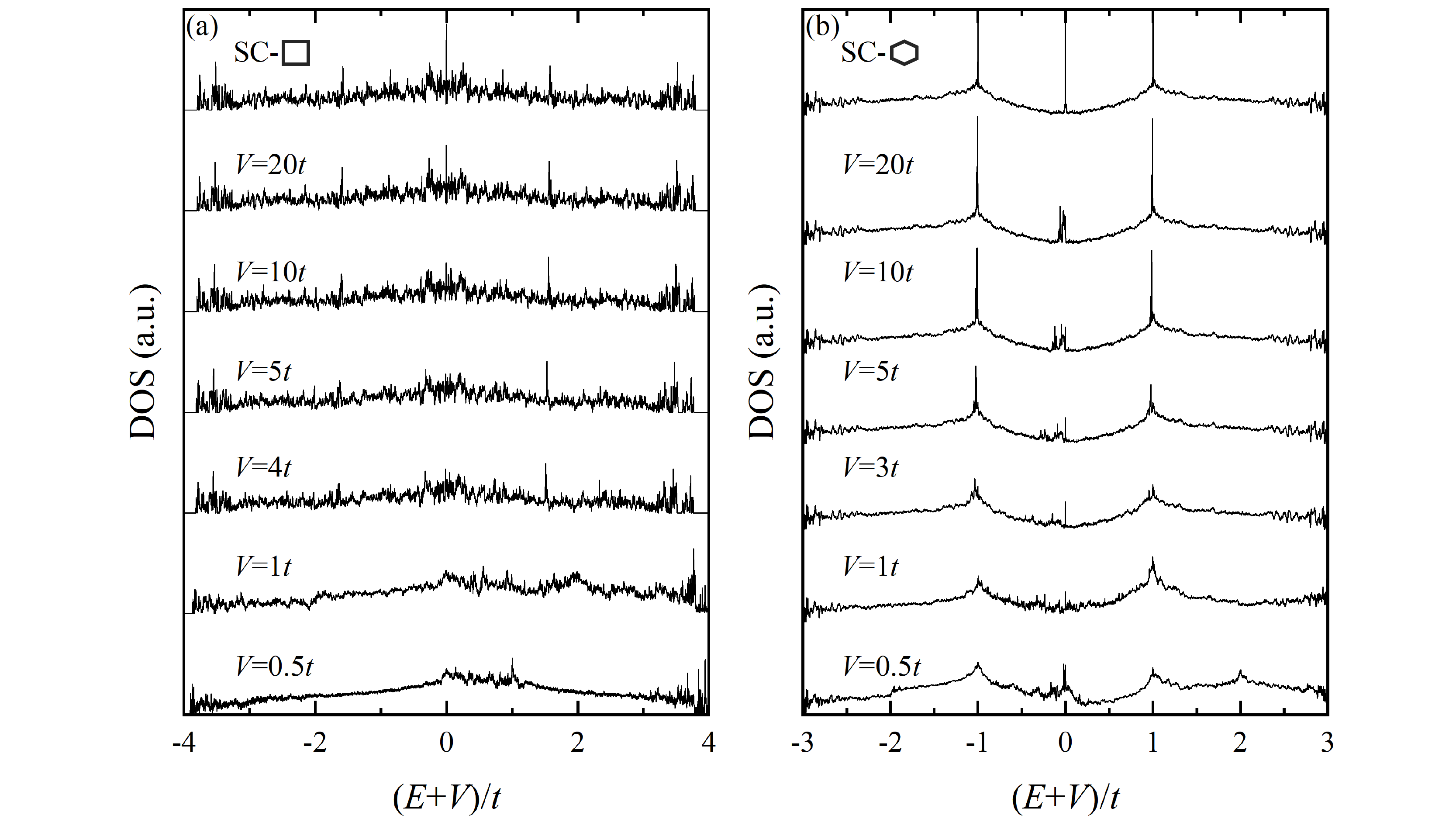}
\caption{(a) The comparison between converged DOS of SC-$\square$ with $I=5$, $W=486$ and corresponding effective ones with different on-site potential $V$. (b) The comparison between converged DOS of SC-$\varhexagon$ with $I=4$, $W=486$ and corresponding effective ones with different on-site potential $V$.}
\label{DOS_all}
\end{figure*}

\section{RESULTS AND DISCUSSION}\label{sec:result}
\subsection{Density of states}
The DOS of the SC-$\square$ with different $I$ as a function of energy is shown in Fig. \ref{DOS_converged}(a).
We change the number of iterations $I$ from $3$ to $6$, and it is clear that the DOS almost converges at $I=5$. Focusing on the SC-$\square$ with $I=5$, the fluctuation of the DOS appears featureless, and edge states result in a central peak at $E/t=0$.
To study the effective SC-$\square$, we first construct a corresponding complete rectangular model, then add opposite on-site potentials onto different areas as mentioned in Fig.~\ref{sample}(b). As we know that if $V$ is infinite, the two areas, with $V$ or $-V$, will be separated completely in the energy spectrum. If $V$ is finite but large enough, we expect that the states in the two areas can still be separated effectively. Indeed, for a square lattice with only nearest-neighbor hoppings ($t_{ij}=t$), as described by the Hamiltonian in Eq.~\ref{equation1}, the energy range will be $[-4t,4t]$, therefore, if $V\geq4t$, the eigenstates in the two areas will be separated in the energy spectrum. As an example, the calculated DOS of the effective SC-$\square$ with $V=4t$ is shown in Fig.~\ref{DOS_converged}(b). The DOS of the effective SC-$\square$ is divided into two parts with a gap between them. The left part ranging from $E/t=-8$ to $0$ is contributed by Area I with the negative on-site potential $-V$ added, while the right part ranging from $E/t=0$ to $8$ is contributed by Area II with the positive on-site potential $V$ added. The DOS of the SC-$\varhexagon$ with different $I$ is shown in Fig.~\ref{DOS_converged}(c) and the results show that the DOS almost converges at $I=4$. Three peaks are found at $E/t=0$, and $E/t=\pm{1}$, respectively. The central peak at $E/t=0$ originates from the edge states localized along with the zigzag terminations over the sample, and the peaks around $E/t=\pm{1}$ are derived from the van-Hove singularities, similar as these in pristine graphene.
If we apply different on-site potentials in a honeycomb lattice to form the effective SC-$\varhexagon$, we see similar effects as in the effective SC-$\square$ that the energy eigenstates are separated into two groups. If $2V$ is not smaller than the width of the energy range ($6t$) of a pristine honeycomb lattice described by Eq.~\ref{equation1}, the energy gap between these two groups in an effective SC increases as increasing $V$. Inversely, the gap decreases as $V$ decreases, and the two groups overlap when $V<4t$ for the effective SC-$\square$ and $V<3t$ for the effective SC-$\varhexagon$. For effective SC, the separation of the DOS can be explained by the competition between on-site potential and hopping. Actually, on-site potential with opposite signs in these two regions leads to a $2V$ potential barrier between them. If hopping $t$ is much smaller than $2V$, the electron would be unlikely to hop from Area I to Area II, and vice versa. 
In the rest of the paper, most calculations are performed for the SC-$\square$ with $I=5$, the SC-$\varhexagon$ with $I=4$, and their corresponding effective SCs, as these systems are large enough to have converged DOS. We want to mention that, although the converged iterations for SC-$\square$ and SC-$\varhexagon$ are different, their sites are in the same order. Indeed, the number of sites of the effective SC-$\square$ with $I=5$ is 236196, and the number of sites of the effective SC-$\varhexagon$ with $I=4$ is 205024.

As we see that finite electric fields with $V=4t$ for effective SC-$\square$ and $V=3t$ for effective SC-$\varhexagon$ are large enough to separate the states in Area I and II, but it remains unclear whether the states in Area I are the same as those in the real SCs or not. Thus, we calculate the DOS of effective SCs and compare them to the corresponding real SCs. We also change the value of $V$ to study the converging behavior of the states in effective SCs. As the states in the effective fractal are created after applying uniform on-site potential $-V$ in Area I, the center of the energy spectrum of these states have an energy shift $-V$ respect to the neutrality point ($E=0$) of the original lattice, and therefore, the comparison of the DOS should be based on the spectrum with this energy shift.
In Fig.~\ref{DOS_all}, after applying the energy shift, we see clearly that the DOS of an effective SC approaches to the real SC with the increment of $V$. The difference between the real and effective SCs is mainly attributed to the states at the center of the spectrum. In Fig.~\ref{DOS_all}(a), we see that, except for the central peak, the DOS of the effective SC-$\square$ with $V=5t$ is basically the same as the one of the SC-$\square$. However, the central peak does not appear until $V=10t$, suggesting that the states at the central peak are more difficult to be reproduced comparing to the other states.
In Fig.~\ref{DOS_all}(b), similar behavior is found in the DOS of the effective SC-$\varhexagon$. Interestingly, three van Hove singularities at $0t$ and $\pm1t$ are not consistent with those of the SC-$\varhexagon$ even with $V=20t$, while the other part of DOS has been converged when $V=3t$.
Based on these results, we conclude that the reproduction of exactly the same spectrum of the DOS over the whole energy range is difficult and requires a very large external field $V$. All the other states which are not around the van Hove singularities are easier to be reproduced in the energy spectrum with a much smaller $V$. 

\begin{figure*}[tbp]
\centering
\includegraphics[width=1.0\textwidth]{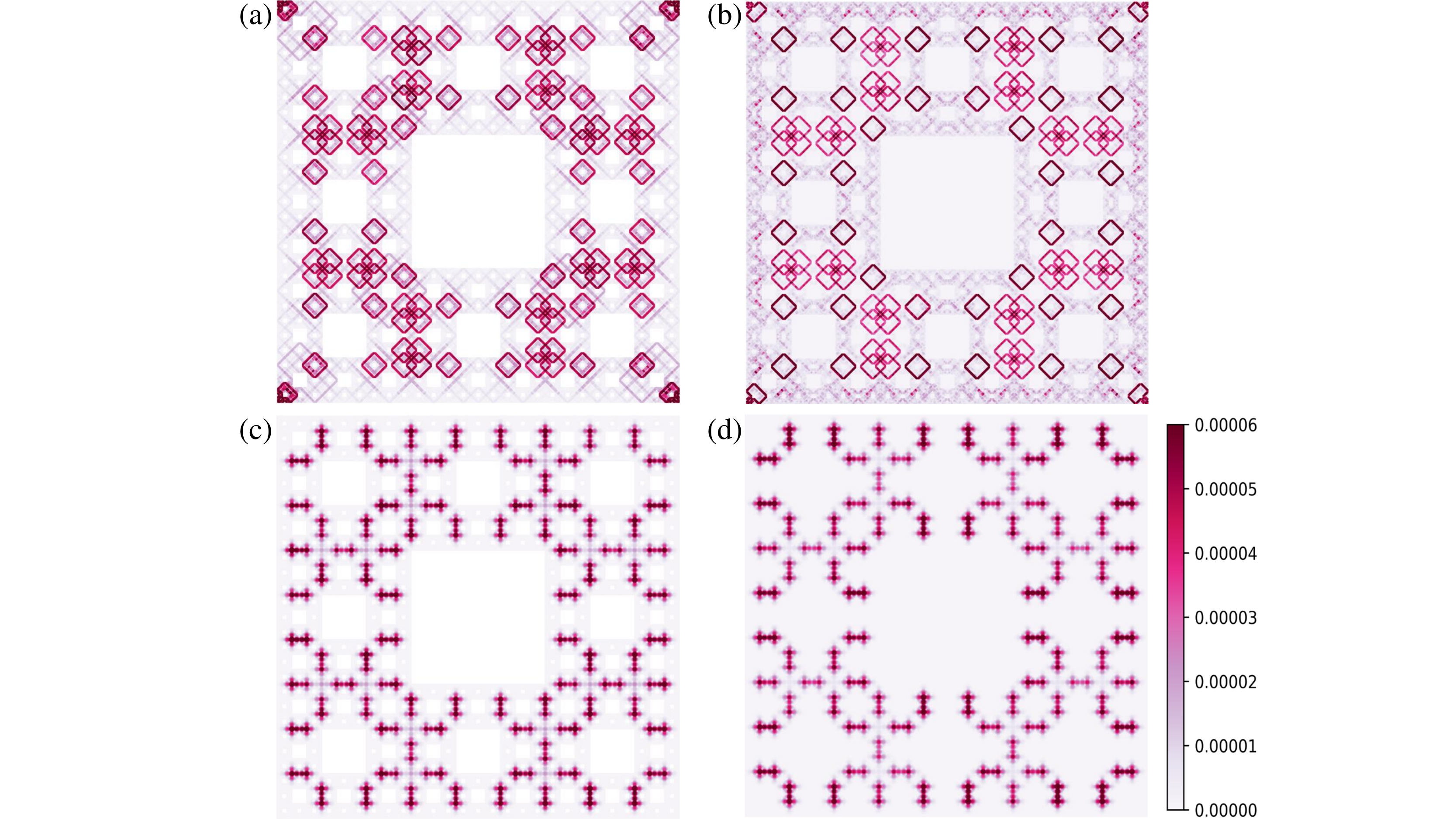}
\caption{(a,c) The real-space distribution of the quasi-eigenstates of the SC-$\square$ with $I=5$, $W=486$ at the central peak $E/t=0$ and the edge peak $E/t=-3.79$. For comparison, we show the real-space distribution of the quasi-eigenstates of the effective SC-$\square$ with $V=3t$ at the corresponding energy in (b,d).}
\label{quasi_squareSC}
\end{figure*}

\begin{figure*}[tbp]
\centering
\includegraphics[width=1.0\textwidth]{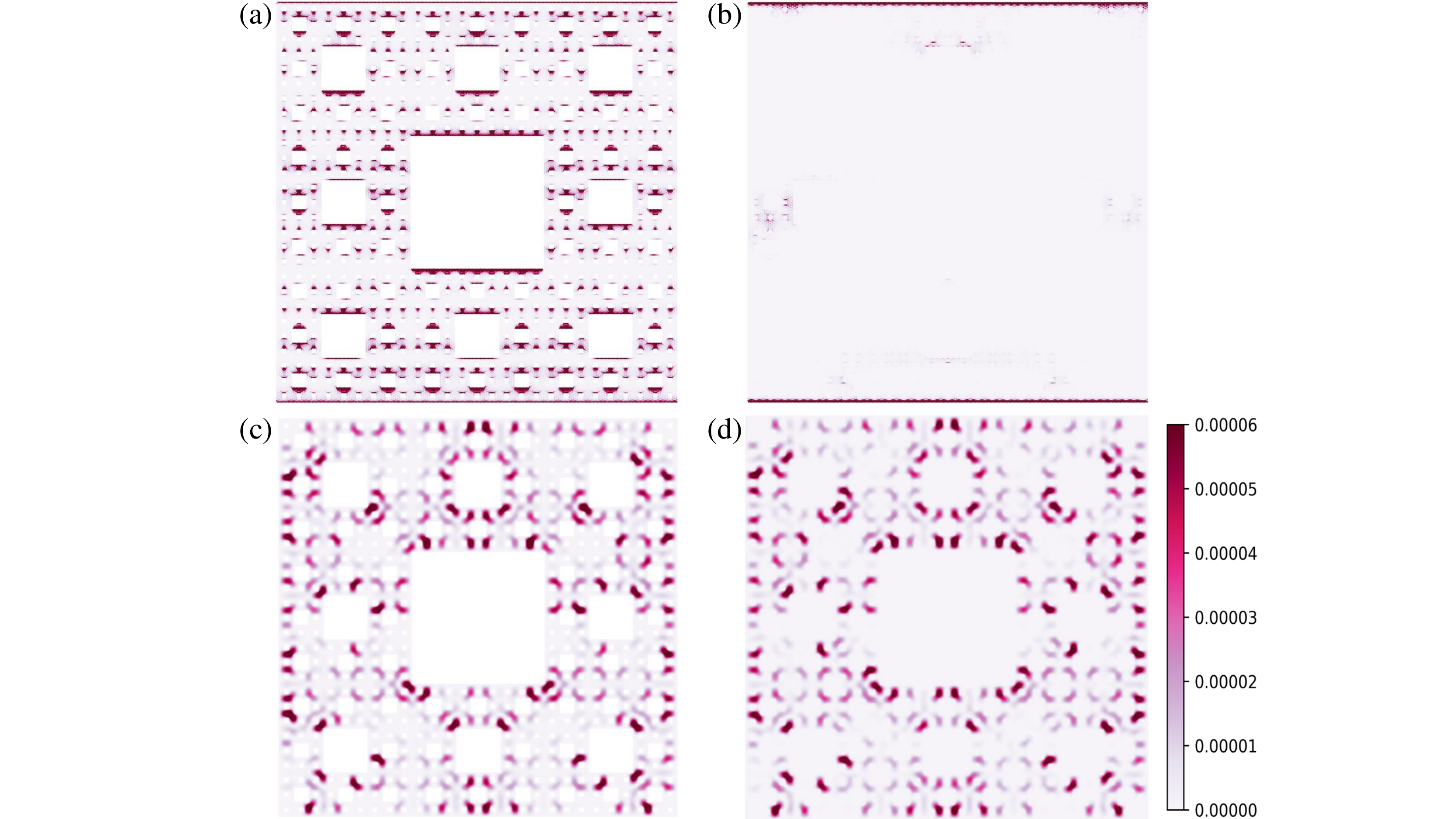}
\caption{(a,c) The real-space distribution of the quasi-eigenstates of the SC-$\varhexagon$ with $I=4$, $W=298$ at the central peak $E/t=0$ and the edge peak $E/t=-2.95$. For comparison, we show the real-space distribution of the quasi-eigenstates of the effective SC-$\varhexagon$ with $V=3t$ at the corresponding energy in (b,d).}
\label{quasi_hexagonalSC}
\end{figure*}

\subsection{Quasi-eigenstates}
In Fig.~\ref{quasi_squareSC}(a), we show the real-space distribution of the quasi-eigenstates at $E/t=0$ of the SC-$\square$. In addition to the central peak, we also show the quasi-eigenstates at $E/t=-3.79$ which is away from the Fermi level (Fig.~\ref{quasi_squareSC}(c)). In Fig.~\ref{quasi_squareSC}(b,d), the real-space distribution of corresponding quasi-eigenstates of the effective SC-$\square$ are plotted. The comparison of the amplitude distributions shows that the quasi-eigenstates of the effective SC-$\square$ with $V=3t$ are in good agreement with those of the SC-$\square$, suggesting that not only the energy spectrum but also the individual states in this effective SC-$\square$ are very similar to real SC-$\square$. As for SC-$\varhexagon$, we also study the effect of the on-site potential on quasi-eigenstates by comparing the calculated quasi-eigenstates of the SC-$\varhexagon$ and the effective one. As shown in Fig.~\ref{quasi_hexagonalSC}(a), the quasi-eigenstate at the central peak with $E/t=0$ originates from the edge states caused by the zigzag edge termination. However, the corresponding quasi-eigenstate of the effective SC-$\varhexagon$ with $V=3t$ is mainly localized at the top and bottom zigzag edges of the whole sample (see Fig.~\ref{quasi_hexagonalSC}(b)). This also explains why the central peak in the DOS of SC-$\varhexagon$ is difficult to be reproduced in the effective SC-$\varhexagon$ by applying finite on-site potentials. In fact, the zero models in the honeycomb lattice are associated with the sites where the sub-lattice symmetry is broken, while in the effective SC-$\varhexagon$, only the top and bottom zigzag edges full fill this condition. The finite potential difference at the boundary of effective "zigzag edges" in the honeycomb lattice is not enough to break the sub-lattice symmetry or create highly localized zero models as those in the real zigzag edges. However, if we are looking at states away from the "neutrality" point, the quasi-eigenstates of the effective SC-$\varhexagon$ are in good agreement with those of SC-$\varhexagon$, see the comparison of the states in Fig.~\ref{quasi_hexagonalSC}(c,d).

In order to get more quantitative information, we sum over the amplitudes of the normalized quasi-eigenstate in Area I, which can be a measure of the total distribution. We call this quantity as the occupation number $R$, and if $R$ is 1, then all the states at the corresponding energy are distributed only in Area I, meaning that an electron with this energy is completely confined in the 'fractal' region. If $R$ is close to 1, then the electron is mostly confined in Area I, with only a small extension to space outside the 'fractal' region. Of course, on the contrary, if $R$ is 0, then the electron has no access to any site in Area I.
In Table. \ref{table1}, we show the occupation number $R$ of some quasi-eigenstates in effective SC-$\square$ with $V=0.5t$ and $3t$. Here, these states are randomly selected from the energy region belonging to the 'fractal' part of the spectrum, i.e., within the range of $[E_{b}-V, E_{b}+V]$, where $E_{b}$ are the lowest energy eigenvalues of the pristine square lattice. From the calculated results, we see that with $V=3t$, the occupation number $R$ is about $98\%-99\%$ for all considered states, indicating that these electrons are indeed confined in the fractal geometry. Surprisingly, even for a small value of $V=0.5t$, although the energy spectrum shown in the DOS is completely different from the real fractal, the occupation number of quasi-eigenstates can still reach $90\%$, much larger than we expect. For honeycomb lattice (shown in Table. \ref{table2}), with a similar analysis we see even better confinement of electrons, and the occupation number $R$ can reach larger values comparing to the square lattice with the same $V$. For example, with $V =0.5t$, calculated $R$ in effective SC-$\varhexagon$ are all larger than $94\%$, indicating that electrons in the honeycomb lattice are more easily to be confined in a fractal geometry comparing to the square lattice. Our calculations of the occupation number $R$ in effective fractals indicate that to confine an electron effectively in a fractal dimension, the applied on-site potentials could be much smaller than the values one expected.


\begin{figure*}[tbp]
\includegraphics[width=0.92\textwidth]{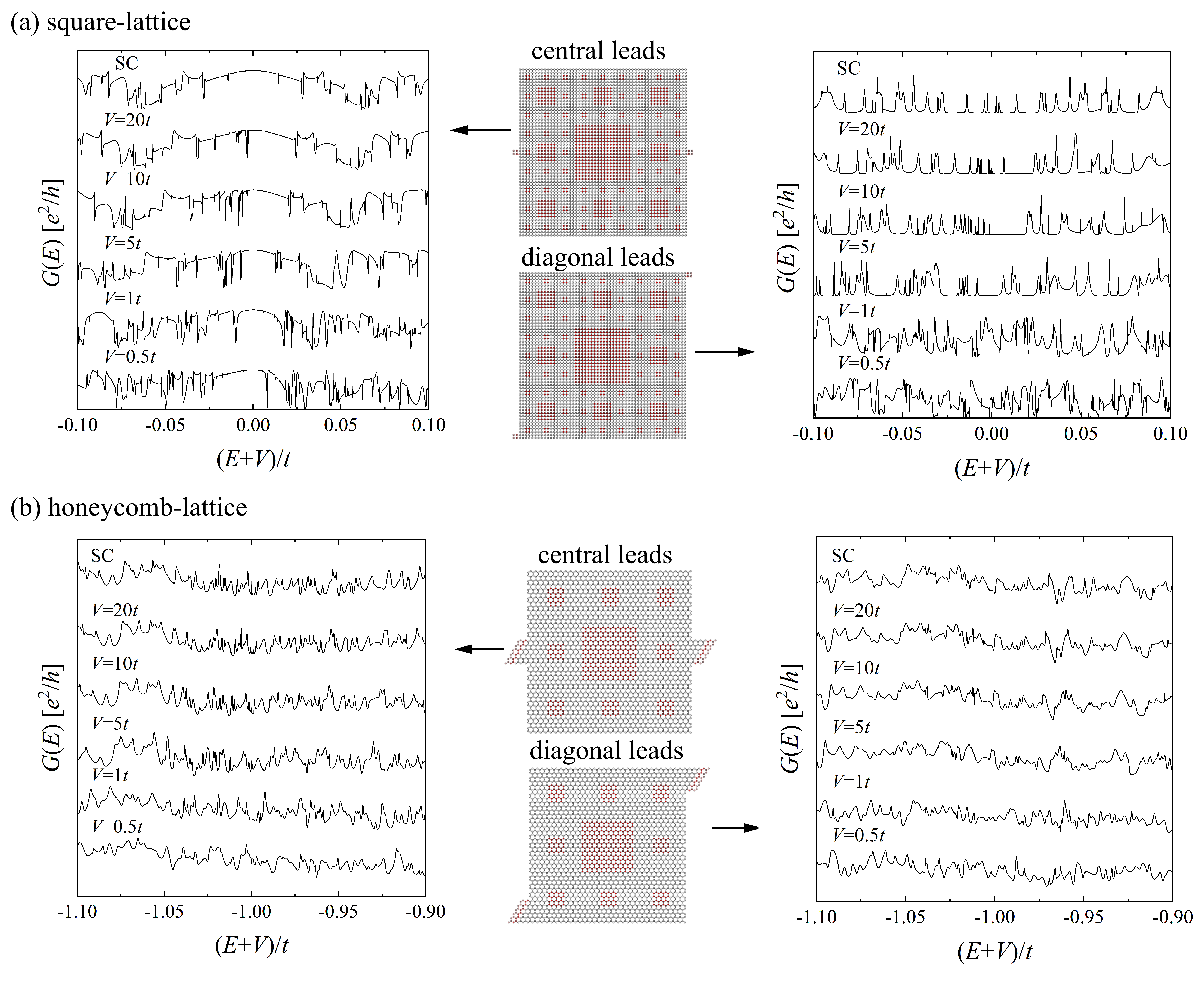}
\caption{Energy dependence of the conductance $G(E)$ (in units of $e^{2}/h$ ) of the (real and effective) SC-$\square$ (a) and SC-$\varhexagon$ (b) with different $V$. $I=3$ and $W=54$ for the (real and effective) SC-$\square$ and $I=2$, $W=33$ for the (real and effective) SC-$\varhexagon$. Left and right panels refer to central lead and diagonal lead configurations respectively. Here, the converges of the conductance fluctuations in different geometries and lead configurations are presented, and we see that for similar lead configurations, the coverage of the spectrum in SC-$\varhexagon$ is faster than SC-$\square$. It is worth to mention that we show here results with relatively smaller samples because the conductance fluctuation of a large system is too dense for visual comparison.
}
\label{conductance}
\end{figure*}

\begin{table}[H]
    \centering
    \caption{The occupation number $R$ of the effective SC-$\square$ with iteration $I=5$, width $W=486$. Two cases where on-site potential $V=0.5t$ (left panel) and $V=3t$ (right panel) are shown.}
    \begin{tabular}{p{2cm}<{\centering}p{2cm}<{\centering}|p{2cm}<{\centering}p{2cm}<\centering}
    \hline
    \multicolumn{2}{p{4cm}<\centering|}{$V=0.5t$}& \multicolumn{2}{p{4cm}<\centering}{$V=3t$}\\
    \hline
    $E/t$&$R$($\%$)&$E/t$&$R$($\%$)\\
    \hline
    -4.399&95.2&-6.824&99.5\\
    -4.258&92.8&-5.648&99.0\\
    -4.088&92.7&-4.647&98.9\\
    -3.832&90.0&-3.023&99.5\\
    -3.650&91.4&-1.512&98.2\\    
    \hline
    \end{tabular}
    \label{table1}
\end{table}

\begin{table}[H]
    \centering
    \caption{The occupation number $R$ of the effective SC-$\varhexagon$ with iteration $I=4$, width $W=298$. Two cases where on-site potential $V=0.5t$ (left panel) and $V=3t$ (right panel) are shown.}
    \begin{tabular}{p{2cm}<{\centering}p{2cm}<{\centering}|p{2cm}<{\centering}p{2cm}<\centering}
    \hline
    \multicolumn{2}{p{4cm}<\centering|}{$V=0.5t$}& \multicolumn{2}{p{4cm}<\centering}{$V=3t$}\\
    \hline
    $E/t$&$R$($\%$)&$E/t$&$R$($\%$)\\
    \hline
    -3.461&99.6&-5.574&99.9\\
    -3.235&98.9&-4.343&99.8\\
    -3.008&97.4&-3.051&99.0\\
    -2.808&97.1&-1.909&99.7\\
    -2.607&94.4&-0.504&99.6\\
    \hline
    \end{tabular}
    \label{table2}
\end{table}

\subsection{Quantum fluctuations of conductance}
Our previous quantum transport calculations of some fractals in Ref.~\cite{transport} show a very interesting and unique character of electrons confined in a fractional space, namely, there is a high correlation between the quantum conductance fluctuations (CFs) and the geometry dimension of the fractal.
Fractal CFs have been verified in many system, such as chaotic systems \cite{ketzmerick1996fractal,sachrajda1998fractal,kotimaki2013fractal}, quantum billiard \cite{crook2003imaging}, gold nanowires \cite{hegger1996fractal}, and diffusive and ballistic semiconductor devices \cite{marlow2006unified}. In Ref. \cite{transport}, quantum CFs of a fractal is characterized by the dimension of its conductance spectrum, which is calculated by the standard box-counting (BC) algorithm \cite{guarneri2001fractal}. For a fractal with an infinite ramification number (in the limit of infinite size) \cite{gefen1984phase}, such as Sierpinski carpet, the box-counting dimension of the quantum CFs is found to be close to the geometry dimension of the fractal \cite{transport}. However, for a fractal with a finite ramification number, there is no such kind of connection between the two types of dimensions. As our main purpose of the current paper is to find how to create an effective fractional dimension by using finite on-site potentials, it is interesting to check whether a similar correlation exists in these effective fractals.

First, in order to study the effect of the on-site potential $V$ on transport properties of the effective SC, we calculated the energy-dependent conductance $G_{ab}(E)$ for a configuration with central leads (two leads are attached to the center of the left and right sides of the scattering region) and one with diagonal leads (one lead attached to the bottom of the left side and the other lead attached to the top of the right side of the scattering region) using Eq.~\ref{equation4} as implemented with KWANT \cite{groth2014kwant}. In the absence of a magnetic field, the quantum conductance can reach the maximum by changing the number of leads, leads positions, and leads width \cite{transport}. 

The numerical results of $G_{ab}(E)$ with different lead configurations are obtained for the effective SC-$\square$ (Fig.~\ref{conductance}(a)) and the effective SC-$\varhexagon$ (Fig.~\ref{conductance}(b)). As the whole spectrum contains many fluctuations which make the curve quite noisy, we show here only parts of the spectrum in order to compare it more clearly to the result of the corresponding real fractal. Here, similar to Fig.~\ref{DOS_all}, an energy shift is introduced for the comparison of the spectrum due to the applied on-site potential. We notice that with this energy shift, the spectrum of conductance fluctuation in effective fractals is similar to the one shown in the real fractal, and the agreement is better if $V$ is larger. Furthermore, with the same amplitude of $V$, the agreement in the effective fractals based on the honeycomb lattice is better than these based on the square lattice (see the difference converging speeds shown in Fig. \ref{conductance}(a) and (b)). This indicates that it is relatively more easily to create an effective fractal by applying on-site potentials in the honeycomb lattice than the square lattice. We also want to mention that, there is an extra shift of the conductance spectrum which is not captured by the simple shift of $E=-V$. If $V$ is larger, this extra energy shift is smaller, indicating that the origin of this extra shift could be due to the finite value of $V$. A complete understanding needs more analytical works that are beyond the scope of this paper.

\begin{figure*}[tbp]
\centering
\includegraphics[width=16cm]{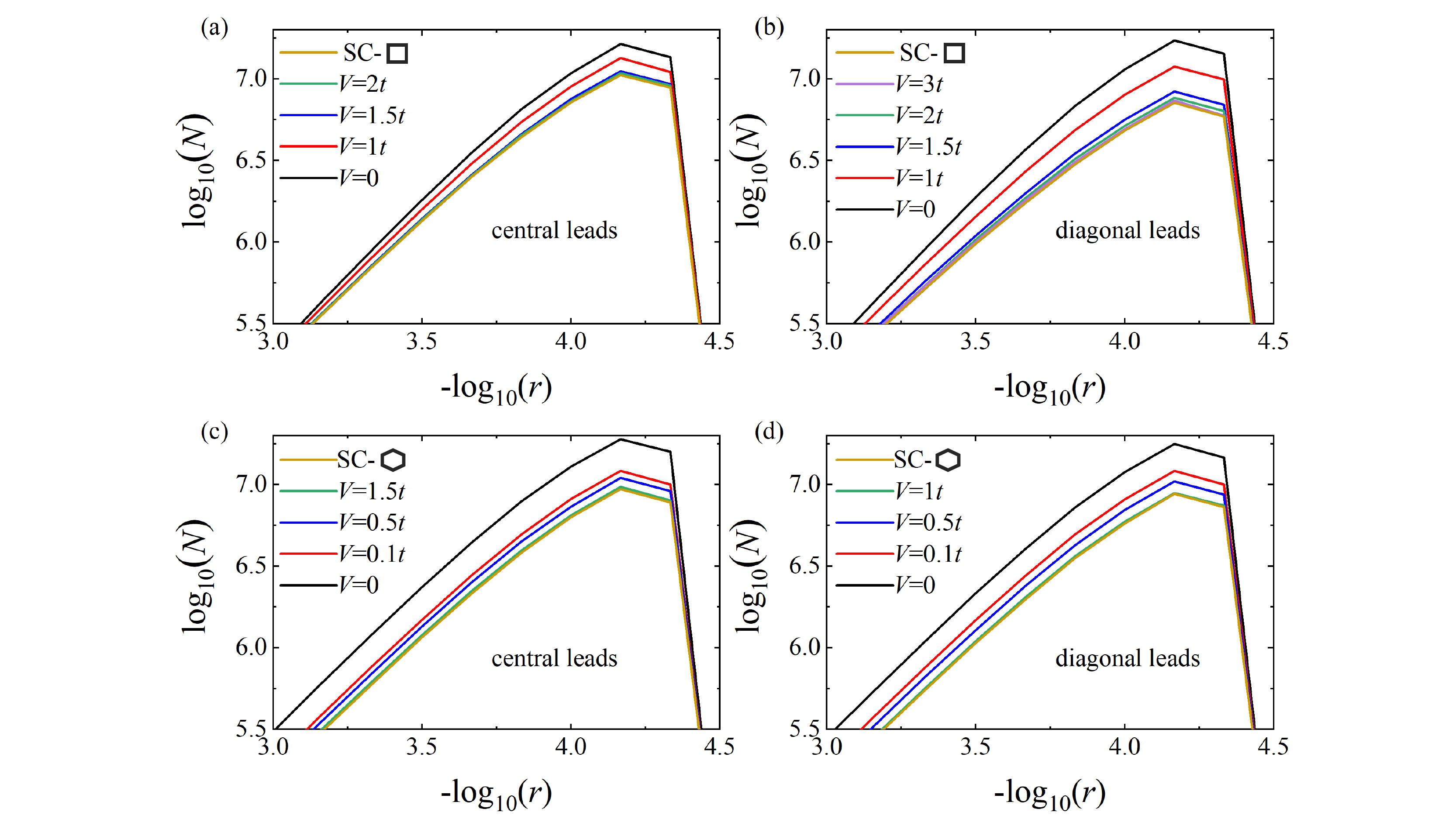}
\caption{BC algorithm analysis of the CFs for SCs. (a,b) BC dimension of the SC-$\square$ with $I=5$, $W=486$ and the corresponding effective ones with different $V$. (c,d) BC dimension of the SC-$\varhexagon$ with $I=4$, $W=298$ and the corresponding effective ones with different $V$. Data in (a,c) and (b,d) refer to central lead and diagonal lead positions respectively.}
\label{BC_dimension}
\end{figure*}

The spectra of the quantum conductance fluctuations shown in Fig. \ref{conductance} are indeed also fractal, and their dimensions can be calculated numerically by using the box-counting algorithm \cite{guarneri2001fractal}. In particular, here we count the number $(N)$ of squares with different size $r$ which continuously and completely cover the graph of conductance ($G(E)$). 
For large values of $r$, the squares are too large to distinguish the features of the graph and $N$ grows slowly as $r$ decreases. Once $r$ decreases to a small enough value where the squares resolve every single point of the raw data of $G(E)$, $N$ is expected to saturate to the number of points in the raw data. Actually, there is an intermediate $r$ called the 'scaling region' where scaling is linear in a log-log plot. Furthermore, the BC dimension of the CFs is determined by the slope $d$ \cite{transport}.
For large $V$, as the graph $G(E)$ of the effective fractals are very close to the real one, it is not surprising that the BC dimensions of these effective fractals agree very well with their geometry dimension, similar as these found in Ref. \cite{transport}. In fact, this agreement converges even faster than other electronic properties considered in previous sections, namely, the density of states and quasi-eigenstates. For example, as our numerical results show in Fig. \ref{BC_dimension}(a) and (b), the BC dimension of the effective SC-$\square$ converges to the value of $V=2t$ for central lead configurations, and $V=3t$ for diagonal lead configurations, respectively. For the effective SC-$\varhexagon$, the agreement occurs earlier with smaller $V$, as one can see from Fig. \ref{BC_dimension}(c) and (d), the cases with $V=1.5t$ for the central lead configurations and $V=t$ for the diagonal lead configurations are enough to reproduce the same BC dimension as in the real fractal.

\section{SUMMARY}\label{sec:conclusion}
In this work, we proposed a way to construct effective fractals from two-dimensional crystals without breaking the atomic structure. We applied an external electric field with fractal geometry, and parts of the electrons are confined effectively in a fractional dimension, and their electronic properties are very similar to these in the corresponding real fractal. To study the effect of the external field quantitatively, we performed calculations of the density of states, quasi-eigenstates, and conductance of the effective SC with various on-site potential $V$ and compare them with real SC. Although a perfect reproduction of fractal requires quite large $V$, we can still find many electronic states in the effective SC which are very similar to these in the SC even with small $V$. Furthermore, the box-counting dimension of the quantum conductance fluctuations in the effective SC converges to the value of SC with much smaller $V$ comparing to other calculated properties such as the density of states and quasi-eigenstates.

Our numerical results indicate that, depending on the properties measured in the experiments, the critical electric field to build an effective fractal will also be different. To calculate the box-counting dimension of quantum conductance fluctuations turns out to be the easiest way to indicate the underlying geometry dimension, but from a practical point of view, the measurement of the whole spectrum of the conductance requires deep doping of the system which may be too difficult or even unreachable currently. In fact, as we show in the comparison of the density of states and quasi-eigenstates, although the states at the center of the energy spectrum are not reproducible with small on-site potentials, the other states can be much easier to be formed with relatively small on-site potentials. The difficulty to reproduce the states at or around the center of the energy spectrum is because these states are mainly localized at the sharp edges of the holes. For example, the midgap states in SC-$\varhexagon$ are due to the breaking of the sub-lattice symmetry at the zigzag edges, and therefore can not be reproduced by the finite on-site potential difference at the boundary between two areas. Thus, we suggest that one should probe the states which are not at the center of the spectrum in order to check whether they are confined effectively in a fractional geometry or not. The comparison of the results with different underlying lattices also suggests that the graphene-like systems, such as graphene, h-BN, MoS$_{2}$ and other transition metal dichalcogenides which have underlying honeycomb lattice, are much easier to form an effective fractional dimension with a small electric field.

As we discussed in the main text, in order to create an effective fractional dimension, one can also apply the electric field only on one area of the lattice, either on the original hole region or the area with fractal geometry. Different ways to apply the field will change the whole spectrum accordingly, either shift all the states with constant energy, or just exchange the states respect to the center of the spectrum. All the electronic properties of corresponding states remain the same. Thus in practice, it would be more convenient to change only the potentials in one region. For the Sierpinski carpet considered in this paper, its geometry dimension is close to 2, the sites in the $hole$ region have less numbers comparing to the $fractal$ region, therefore it is much easier to manipulate the $hole$ region. For some other fractals, one may choose to control oppositely the $fractal$ region.

As a conclusion, we propose a way to confine electrons in an effective fractional dimension by applying an external electric field with fractal geometry. Our work paves a new way to realize fractals from top to bottom without destroying the atomic structure of the underlying lattice. One can also control the electric field to make the whole process reversible, i.e., to create and destroy the effective fractional dimension by change the electric field. We believe our work will motivate more experimental and theoretical studies of fractals in real systems.

\begin{acknowledgements}
This work is supported by the National Nature Science Foundation of China (Grant No. 11774269). Numerical calculations presented in this paper have been performed on a supercomputing system in the Supercomputing Center of Wuhan University.
\end{acknowledgements}

\bibliographystyle{apsrev4-1}
\bibliography{ref}

\begin{thebibliography}{59}%
\makeatletter
\providecommand \@ifxundefined [1]{%
 \@ifx{#1\undefined}
}%
\providecommand \@ifnum [1]{%
 \ifnum #1\expandafter \@firstoftwo
 \else \expandafter \@secondoftwo
 \fi
}%
\providecommand \@ifx [1]{%
 \ifx #1\expandafter \@firstoftwo
 \else \expandafter \@secondoftwo
 \fi
}%
\providecommand \natexlab [1]{#1}%
\providecommand \enquote  [1]{``#1''}%
\providecommand \bibnamefont  [1]{#1}%
\providecommand \bibfnamefont [1]{#1}%
\providecommand \citenamefont [1]{#1}%
\providecommand \href@noop [0]{\@secondoftwo}%
\providecommand \href [0]{\begingroup \@sanitize@url \@href}%
\providecommand \@href[1]{\@@startlink{#1}\@@href}%
\providecommand \@@href[1]{\endgroup#1\@@endlink}%
\providecommand \@sanitize@url [0]{\catcode `\\12\catcode `\$12\catcode
  `\&12\catcode `\#12\catcode `\^12\catcode `\_12\catcode `\%12\relax}%
\providecommand \@@startlink[1]{}%
\providecommand \@@endlink[0]{}%
\providecommand \url  [0]{\begingroup\@sanitize@url \@url }%
\providecommand \@url [1]{\endgroup\@href {#1}{\urlprefix }}%
\providecommand \urlprefix  [0]{URL }%
\providecommand \Eprint [0]{\href }%
\providecommand \doibase [0]{http://dx.doi.org/}%
\providecommand \selectlanguage [0]{\@gobble}%
\providecommand \bibinfo  [0]{\@secondoftwo}%
\providecommand \bibfield  [0]{\@secondoftwo}%
\providecommand \translation [1]{[#1]}%
\providecommand \BibitemOpen [0]{}%
\providecommand \bibitemStop [0]{}%
\providecommand \bibitemNoStop [0]{.\EOS\space}%
\providecommand \EOS [0]{\spacefactor3000\relax}%
\providecommand \BibitemShut  [1]{\csname bibitem#1\endcsname}%
\let\auto@bib@innerbib\@empty
\bibitem [{\citenamefont {Falconer}(1985)}]{geometryoffractal}%
  \BibitemOpen
  \bibfield  {author} {\bibinfo {author} {\bibfnamefont {K.~J.}\ \bibnamefont
  {Falconer}},\ }\href@noop {} {\emph {\bibinfo {title} {The Geometry of
  Fractal Sets}}}\ (\bibinfo  {publisher} {Cambridge University Press,
  Cambridge},\ \bibinfo {year} {1985})\BibitemShut {NoStop}%
\bibitem [{\citenamefont {Feder}(2013)}]{feder2013fractals}%
  \BibitemOpen
  \bibfield  {author} {\bibinfo {author} {\bibfnamefont {J.}~\bibnamefont
  {Feder}},\ }\href@noop {} {\emph {\bibinfo {title} {Fractals}}}\ (\bibinfo
  {publisher} {Springer Science \& Business Media},\ \bibinfo {year}
  {2013})\BibitemShut {NoStop}%
\bibitem [{\citenamefont {Gomes}\ \emph {et~al.}(2012)\citenamefont {Gomes},
  \citenamefont {Mar}, \citenamefont {Ko}, \citenamefont {Guinea},\ and\
  \citenamefont {Manoharan}}]{gomes2012designer}%
  \BibitemOpen
  \bibfield  {author} {\bibinfo {author} {\bibfnamefont {K.~K.}\ \bibnamefont
  {Gomes}}, \bibinfo {author} {\bibfnamefont {W.}~\bibnamefont {Mar}}, \bibinfo
  {author} {\bibfnamefont {W.}~\bibnamefont {Ko}}, \bibinfo {author}
  {\bibfnamefont {F.}~\bibnamefont {Guinea}}, \ and\ \bibinfo {author}
  {\bibfnamefont {H.~C.}\ \bibnamefont {Manoharan}},\ }\href@noop {} {\bibfield
   {journal} {\bibinfo  {journal} {Nature}\ }\textbf {\bibinfo {volume}
  {483}},\ \bibinfo {pages} {306} (\bibinfo {year} {2012})}\BibitemShut
  {NoStop}%
\bibitem [{\citenamefont {Polini}\ \emph {et~al.}(2013)\citenamefont {Polini},
  \citenamefont {Guinea}, \citenamefont {Lewenstein}, \citenamefont
  {Manoharan},\ and\ \citenamefont {Pellegrini}}]{polini2013artificial}%
  \BibitemOpen
  \bibfield  {author} {\bibinfo {author} {\bibfnamefont {M.}~\bibnamefont
  {Polini}}, \bibinfo {author} {\bibfnamefont {F.}~\bibnamefont {Guinea}},
  \bibinfo {author} {\bibfnamefont {M.}~\bibnamefont {Lewenstein}}, \bibinfo
  {author} {\bibfnamefont {H.~C.}\ \bibnamefont {Manoharan}}, \ and\ \bibinfo
  {author} {\bibfnamefont {V.}~\bibnamefont {Pellegrini}},\ }\href@noop {}
  {\bibfield  {journal} {\bibinfo  {journal} {Nature Nanotech}\ }\textbf
  {\bibinfo {volume} {8}},\ \bibinfo {pages} {625} (\bibinfo {year}
  {2013})}\BibitemShut {NoStop}%
\bibitem [{\citenamefont {Slot}\ \emph {et~al.}(2017)\citenamefont {Slot},
  \citenamefont {Gardenier}, \citenamefont {Jacobse}, \citenamefont {van
  Miert}, \citenamefont {Kempkes}, \citenamefont {Zevenhuizen}, \citenamefont
  {Smith}, \citenamefont {Vanmaekelbergh},\ and\ \citenamefont
  {Swart}}]{slot2017experimental}%
  \BibitemOpen
  \bibfield  {author} {\bibinfo {author} {\bibfnamefont {M.~R.}\ \bibnamefont
  {Slot}}, \bibinfo {author} {\bibfnamefont {T.~S.}\ \bibnamefont {Gardenier}},
  \bibinfo {author} {\bibfnamefont {P.~H.}\ \bibnamefont {Jacobse}}, \bibinfo
  {author} {\bibfnamefont {G.~C.}\ \bibnamefont {van Miert}}, \bibinfo {author}
  {\bibfnamefont {S.~N.}\ \bibnamefont {Kempkes}}, \bibinfo {author}
  {\bibfnamefont {S.~J.}\ \bibnamefont {Zevenhuizen}}, \bibinfo {author}
  {\bibfnamefont {C.~M.}\ \bibnamefont {Smith}}, \bibinfo {author}
  {\bibfnamefont {D.}~\bibnamefont {Vanmaekelbergh}}, \ and\ \bibinfo {author}
  {\bibfnamefont {I.}~\bibnamefont {Swart}},\ }\href@noop {} {\bibfield
  {journal} {\bibinfo  {journal} {Nature Phys.}\ }\textbf {\bibinfo {volume}
  {13}},\ \bibinfo {pages} {672} (\bibinfo {year} {2017})}\BibitemShut
  {NoStop}%
\bibitem [{\citenamefont {Scarabelli}\ \emph {et~al.}(2015)\citenamefont
  {Scarabelli}, \citenamefont {Wang}, \citenamefont {Pinczuk}, \citenamefont
  {Wind}, \citenamefont {Kuznetsova}, \citenamefont {Pfeiffer}, \citenamefont
  {West}, \citenamefont {Gardner}, \citenamefont {Manfra},\ and\ \citenamefont
  {Pellegrini}}]{scarabelli2015fabrication}%
  \BibitemOpen
  \bibfield  {author} {\bibinfo {author} {\bibfnamefont {D.}~\bibnamefont
  {Scarabelli}}, \bibinfo {author} {\bibfnamefont {S.}~\bibnamefont {Wang}},
  \bibinfo {author} {\bibfnamefont {A.}~\bibnamefont {Pinczuk}}, \bibinfo
  {author} {\bibfnamefont {S.~J.}\ \bibnamefont {Wind}}, \bibinfo {author}
  {\bibfnamefont {Y.~Y.}\ \bibnamefont {Kuznetsova}}, \bibinfo {author}
  {\bibfnamefont {L.~N.}\ \bibnamefont {Pfeiffer}}, \bibinfo {author}
  {\bibfnamefont {K.}~\bibnamefont {West}}, \bibinfo {author} {\bibfnamefont
  {G.~C.}\ \bibnamefont {Gardner}}, \bibinfo {author} {\bibfnamefont {M.~J.}\
  \bibnamefont {Manfra}}, \ and\ \bibinfo {author} {\bibfnamefont
  {V.}~\bibnamefont {Pellegrini}},\ }\href@noop {} {\bibfield  {journal}
  {\bibinfo  {journal} {J. Vac. Sci. Technol. B}\ }\textbf {\bibinfo {volume}
  {33}},\ \bibinfo {pages} {06FG03} (\bibinfo {year} {2015})}\BibitemShut
  {NoStop}%
\bibitem [{\citenamefont {De~Simoni}\ \emph {et~al.}(2010)\citenamefont
  {De~Simoni}, \citenamefont {Singha}, \citenamefont {Gibertini}, \citenamefont
  {Karmakar}, \citenamefont {Polini}, \citenamefont {Piazza}, \citenamefont
  {Pfeiffer}, \citenamefont {West}, \citenamefont {Beltram},\ and\
  \citenamefont {Pellegrini}}]{de2010delocalized}%
  \BibitemOpen
  \bibfield  {author} {\bibinfo {author} {\bibfnamefont {G.}~\bibnamefont
  {De~Simoni}}, \bibinfo {author} {\bibfnamefont {A.}~\bibnamefont {Singha}},
  \bibinfo {author} {\bibfnamefont {M.}~\bibnamefont {Gibertini}}, \bibinfo
  {author} {\bibfnamefont {B.}~\bibnamefont {Karmakar}}, \bibinfo {author}
  {\bibfnamefont {M.}~\bibnamefont {Polini}}, \bibinfo {author} {\bibfnamefont
  {V.}~\bibnamefont {Piazza}}, \bibinfo {author} {\bibfnamefont
  {L.}~\bibnamefont {Pfeiffer}}, \bibinfo {author} {\bibfnamefont
  {K.}~\bibnamefont {West}}, \bibinfo {author} {\bibfnamefont {F.}~\bibnamefont
  {Beltram}}, \ and\ \bibinfo {author} {\bibfnamefont {V.}~\bibnamefont
  {Pellegrini}},\ }\href@noop {} {\bibfield  {journal} {\bibinfo  {journal}
  {Appl. Phys. Lett.}\ }\textbf {\bibinfo {volume} {97}},\ \bibinfo {pages}
  {132113} (\bibinfo {year} {2010})}\BibitemShut {NoStop}%
\bibitem [{\citenamefont {Nadvornik}\ \emph {et~al.}(2012)\citenamefont
  {Nadvornik}, \citenamefont {Orlita}, \citenamefont {Goncharuk}, \citenamefont
  {Smr{\v{c}}ka}, \citenamefont {Nov{\'a}k}, \citenamefont {Jurka},
  \citenamefont {Hru{\v{s}}ka}, \citenamefont {V{\`y}born{\`y}}, \citenamefont
  {Wasilewski}, \citenamefont {Potemski} \emph
  {et~al.}}]{nadvornik2012laterally}%
  \BibitemOpen
  \bibfield  {author} {\bibinfo {author} {\bibfnamefont {L.}~\bibnamefont
  {Nadvornik}}, \bibinfo {author} {\bibfnamefont {M.}~\bibnamefont {Orlita}},
  \bibinfo {author} {\bibfnamefont {N.}~\bibnamefont {Goncharuk}}, \bibinfo
  {author} {\bibfnamefont {L.}~\bibnamefont {Smr{\v{c}}ka}}, \bibinfo {author}
  {\bibfnamefont {V.}~\bibnamefont {Nov{\'a}k}}, \bibinfo {author}
  {\bibfnamefont {V.}~\bibnamefont {Jurka}}, \bibinfo {author} {\bibfnamefont
  {K.}~\bibnamefont {Hru{\v{s}}ka}}, \bibinfo {author} {\bibfnamefont
  {Z.}~\bibnamefont {V{\`y}born{\`y}}}, \bibinfo {author} {\bibfnamefont
  {Z.}~\bibnamefont {Wasilewski}}, \bibinfo {author} {\bibfnamefont
  {M.}~\bibnamefont {Potemski}},  \emph {et~al.},\ }\href@noop {} {\bibfield
  {journal} {\bibinfo  {journal} {New J. Phys.}\ }\textbf {\bibinfo {volume}
  {14}},\ \bibinfo {pages} {053002} (\bibinfo {year} {2012})}\BibitemShut
  {NoStop}%
\bibitem [{\citenamefont {Newkome}\ \emph {et~al.}(2006)\citenamefont
  {Newkome}, \citenamefont {Wang}, \citenamefont {Moorefield}, \citenamefont
  {Cho}, \citenamefont {Mohapatra}, \citenamefont {Li}, \citenamefont {Hwang},
  \citenamefont {Lukoyanova}, \citenamefont {Echegoyen}, \citenamefont
  {Palagallo} \emph {et~al.}}]{newkome2006nanoassembly}%
  \BibitemOpen
  \bibfield  {author} {\bibinfo {author} {\bibfnamefont {G.~R.}\ \bibnamefont
  {Newkome}}, \bibinfo {author} {\bibfnamefont {P.}~\bibnamefont {Wang}},
  \bibinfo {author} {\bibfnamefont {C.~N.}\ \bibnamefont {Moorefield}},
  \bibinfo {author} {\bibfnamefont {T.~J.}\ \bibnamefont {Cho}}, \bibinfo
  {author} {\bibfnamefont {P.~P.}\ \bibnamefont {Mohapatra}}, \bibinfo {author}
  {\bibfnamefont {S.}~\bibnamefont {Li}}, \bibinfo {author} {\bibfnamefont
  {S.-H.}\ \bibnamefont {Hwang}}, \bibinfo {author} {\bibfnamefont
  {O.}~\bibnamefont {Lukoyanova}}, \bibinfo {author} {\bibfnamefont
  {L.}~\bibnamefont {Echegoyen}}, \bibinfo {author} {\bibfnamefont {J.~A.}\
  \bibnamefont {Palagallo}},  \emph {et~al.},\ }\href@noop {} {\bibfield
  {journal} {\bibinfo  {journal} {Science}\ }\textbf {\bibinfo {volume}
  {312}},\ \bibinfo {pages} {1782} (\bibinfo {year} {2006})}\BibitemShut
  {NoStop}%
\bibitem [{\citenamefont {Shang}\ \emph {et~al.}(2015)\citenamefont {Shang},
  \citenamefont {Wang}, \citenamefont {Chen}, \citenamefont {Dai},
  \citenamefont {Zhou}, \citenamefont {Kuttner}, \citenamefont {Hilt},
  \citenamefont {Shao}, \citenamefont {Gottfried},\ and\ \citenamefont
  {Wu}}]{shang2015assembling}%
  \BibitemOpen
  \bibfield  {author} {\bibinfo {author} {\bibfnamefont {J.}~\bibnamefont
  {Shang}}, \bibinfo {author} {\bibfnamefont {Y.}~\bibnamefont {Wang}},
  \bibinfo {author} {\bibfnamefont {M.}~\bibnamefont {Chen}}, \bibinfo {author}
  {\bibfnamefont {J.}~\bibnamefont {Dai}}, \bibinfo {author} {\bibfnamefont
  {X.}~\bibnamefont {Zhou}}, \bibinfo {author} {\bibfnamefont {J.}~\bibnamefont
  {Kuttner}}, \bibinfo {author} {\bibfnamefont {G.}~\bibnamefont {Hilt}},
  \bibinfo {author} {\bibfnamefont {X.}~\bibnamefont {Shao}}, \bibinfo {author}
  {\bibfnamefont {J.~M.}\ \bibnamefont {Gottfried}}, \ and\ \bibinfo {author}
  {\bibfnamefont {K.}~\bibnamefont {Wu}},\ }\href@noop {} {\bibfield  {journal}
  {\bibinfo  {journal} {Nature Chem.}\ }\textbf {\bibinfo {volume} {7}},\
  \bibinfo {pages} {389} (\bibinfo {year} {2015})}\BibitemShut {NoStop}%
\bibitem [{\citenamefont {Zhang}\ \emph {et~al.}(2016)\citenamefont {Zhang},
  \citenamefont {Li}, \citenamefont {Liu}, \citenamefont {Gu}, \citenamefont
  {Li}, \citenamefont {Tang}, \citenamefont {Peng}, \citenamefont {Hou},\ and\
  \citenamefont {Wang}}]{zhang2016robust}%
  \BibitemOpen
  \bibfield  {author} {\bibinfo {author} {\bibfnamefont {X.}~\bibnamefont
  {Zhang}}, \bibinfo {author} {\bibfnamefont {N.}~\bibnamefont {Li}}, \bibinfo
  {author} {\bibfnamefont {L.}~\bibnamefont {Liu}}, \bibinfo {author}
  {\bibfnamefont {G.}~\bibnamefont {Gu}}, \bibinfo {author} {\bibfnamefont
  {C.}~\bibnamefont {Li}}, \bibinfo {author} {\bibfnamefont {H.}~\bibnamefont
  {Tang}}, \bibinfo {author} {\bibfnamefont {L.}~\bibnamefont {Peng}}, \bibinfo
  {author} {\bibfnamefont {S.}~\bibnamefont {Hou}}, \ and\ \bibinfo {author}
  {\bibfnamefont {Y.}~\bibnamefont {Wang}},\ }\href@noop {} {\bibfield
  {journal} {\bibinfo  {journal} {ChemComm}\ }\textbf {\bibinfo {volume}
  {52}},\ \bibinfo {pages} {10578} (\bibinfo {year} {2016})}\BibitemShut
  {NoStop}%
\bibitem [{\citenamefont {Kempkes}\ \emph {et~al.}(2019)\citenamefont
  {Kempkes}, \citenamefont {Slot}, \citenamefont {Freeney}, \citenamefont
  {Zevenhuizen}, \citenamefont {Vanmaekelbergh}, \citenamefont {Swart},\ and\
  \citenamefont {Smith}}]{kempkes2019design}%
  \BibitemOpen
  \bibfield  {author} {\bibinfo {author} {\bibfnamefont {S.~N.}\ \bibnamefont
  {Kempkes}}, \bibinfo {author} {\bibfnamefont {M.~R.}\ \bibnamefont {Slot}},
  \bibinfo {author} {\bibfnamefont {S.~E.}\ \bibnamefont {Freeney}}, \bibinfo
  {author} {\bibfnamefont {S.~J.}\ \bibnamefont {Zevenhuizen}}, \bibinfo
  {author} {\bibfnamefont {D.}~\bibnamefont {Vanmaekelbergh}}, \bibinfo
  {author} {\bibfnamefont {I.}~\bibnamefont {Swart}}, \ and\ \bibinfo {author}
  {\bibfnamefont {C.~M.}\ \bibnamefont {Smith}},\ }\href@noop {} {\bibfield
  {journal} {\bibinfo  {journal} {Nature Phys.}\ }\textbf {\bibinfo {volume}
  {15}},\ \bibinfo {pages} {127} (\bibinfo {year} {2019})}\BibitemShut
  {NoStop}%
\bibitem [{\citenamefont {van Veen}\ \emph {et~al.}(2016)\citenamefont {van
  Veen}, \citenamefont {Yuan}, \citenamefont {Katsnelson}, \citenamefont
  {Polini},\ and\ \citenamefont {Tomadin}}]{transport}%
  \BibitemOpen
  \bibfield  {author} {\bibinfo {author} {\bibfnamefont {E.}~\bibnamefont {van
  Veen}}, \bibinfo {author} {\bibfnamefont {S.}~\bibnamefont {Yuan}}, \bibinfo
  {author} {\bibfnamefont {M.~I.}\ \bibnamefont {Katsnelson}}, \bibinfo
  {author} {\bibfnamefont {M.}~\bibnamefont {Polini}}, \ and\ \bibinfo {author}
  {\bibfnamefont {A.}~\bibnamefont {Tomadin}},\ }\href@noop {} {\bibfield
  {journal} {\bibinfo  {journal} {Phys. Rev. B}\ }\textbf {\bibinfo {volume}
  {93}},\ \bibinfo {pages} {115428} (\bibinfo {year} {2016})}\BibitemShut
  {NoStop}%
\bibitem [{\citenamefont {van Veen}\ \emph {et~al.}(2017)\citenamefont {van
  Veen}, \citenamefont {Tomadin}, \citenamefont {Polini}, \citenamefont
  {Katsnelson},\ and\ \citenamefont {Yuan}}]{opticalconductivity}%
  \BibitemOpen
  \bibfield  {author} {\bibinfo {author} {\bibfnamefont {E.}~\bibnamefont {van
  Veen}}, \bibinfo {author} {\bibfnamefont {A.}~\bibnamefont {Tomadin}},
  \bibinfo {author} {\bibfnamefont {M.}~\bibnamefont {Polini}}, \bibinfo
  {author} {\bibfnamefont {M.~I.}\ \bibnamefont {Katsnelson}}, \ and\ \bibinfo
  {author} {\bibfnamefont {S.}~\bibnamefont {Yuan}},\ }\href@noop {} {\bibfield
   {journal} {\bibinfo  {journal} {Phys. Rev. B}\ }\textbf {\bibinfo {volume}
  {96}},\ \bibinfo {pages} {235438} (\bibinfo {year} {2017})}\BibitemShut
  {NoStop}%
\bibitem [{\citenamefont {Han}\ and\ \citenamefont
  {Qiao}(2019)}]{han2019universal}%
  \BibitemOpen
  \bibfield  {author} {\bibinfo {author} {\bibfnamefont {Y.-L.}\ \bibnamefont
  {Han}}\ and\ \bibinfo {author} {\bibfnamefont {Z.-H.}\ \bibnamefont {Qiao}},\
  }\href@noop {} {\bibfield  {journal} {\bibinfo  {journal} {Frontiers of
  Physics}\ }\textbf {\bibinfo {volume} {14}},\ \bibinfo {pages} {63603}
  (\bibinfo {year} {2019})}\BibitemShut {NoStop}%
\bibitem [{\citenamefont {Bouzerar}\ and\ \citenamefont
  {Mayou}(2020)}]{bouzerar2020quantum}%
  \BibitemOpen
  \bibfield  {author} {\bibinfo {author} {\bibfnamefont {G.}~\bibnamefont
  {Bouzerar}}\ and\ \bibinfo {author} {\bibfnamefont {D.}~\bibnamefont
  {Mayou}},\ }\href@noop {} {\bibfield  {journal} {\bibinfo  {journal}
  {Physical Review Research}\ }\textbf {\bibinfo {volume} {2}},\ \bibinfo
  {pages} {033063} (\bibinfo {year} {2020})}\BibitemShut {NoStop}%
\bibitem [{\citenamefont {Iliasov}\ \emph {et~al.}(2020)\citenamefont
  {Iliasov}, \citenamefont {Katsnelson},\ and\ \citenamefont
  {Yuan}}]{iliasov2020hall}%
  \BibitemOpen
  \bibfield  {author} {\bibinfo {author} {\bibfnamefont {A.~A.}\ \bibnamefont
  {Iliasov}}, \bibinfo {author} {\bibfnamefont {M.~I.}\ \bibnamefont
  {Katsnelson}}, \ and\ \bibinfo {author} {\bibfnamefont {S.}~\bibnamefont
  {Yuan}},\ }\href@noop {} {\bibfield  {journal} {\bibinfo  {journal} {Phys.
  Rev. B}\ }\textbf {\bibinfo {volume} {101}},\ \bibinfo {pages} {045413}
  (\bibinfo {year} {2020})}\BibitemShut {NoStop}%
\bibitem [{\citenamefont {Fremling}\ \emph {et~al.}(2020)\citenamefont
  {Fremling}, \citenamefont {van Hooft}, \citenamefont {Smith},\ and\
  \citenamefont {Fritz}}]{fremling2020existence}%
  \BibitemOpen
  \bibfield  {author} {\bibinfo {author} {\bibfnamefont {M.}~\bibnamefont
  {Fremling}}, \bibinfo {author} {\bibfnamefont {M.}~\bibnamefont {van Hooft}},
  \bibinfo {author} {\bibfnamefont {C.~M.}\ \bibnamefont {Smith}}, \ and\
  \bibinfo {author} {\bibfnamefont {L.}~\bibnamefont {Fritz}},\ }\href@noop {}
  {\bibfield  {journal} {\bibinfo  {journal} {Physical Review Research}\
  }\textbf {\bibinfo {volume} {2}},\ \bibinfo {pages} {013044} (\bibinfo {year}
  {2020})}\BibitemShut {NoStop}%
\bibitem [{\citenamefont {Westerhout}\ \emph {et~al.}(2018)\citenamefont
  {Westerhout}, \citenamefont {van Veen}, \citenamefont {Katsnelson},\ and\
  \citenamefont {Yuan}}]{westerhout2018plasmon}%
  \BibitemOpen
  \bibfield  {author} {\bibinfo {author} {\bibfnamefont {T.}~\bibnamefont
  {Westerhout}}, \bibinfo {author} {\bibfnamefont {E.}~\bibnamefont {van
  Veen}}, \bibinfo {author} {\bibfnamefont {M.~I.}\ \bibnamefont {Katsnelson}},
  \ and\ \bibinfo {author} {\bibfnamefont {S.}~\bibnamefont {Yuan}},\
  }\href@noop {} {\bibfield  {journal} {\bibinfo  {journal} {Phys. Rev. B}\
  }\textbf {\bibinfo {volume} {97}},\ \bibinfo {pages} {205434} (\bibinfo
  {year} {2018})}\BibitemShut {NoStop}%
\bibitem [{\citenamefont {Nandy}\ and\ \citenamefont
  {Chakrabarti}(2015)}]{nandy2015engineering}%
  \BibitemOpen
  \bibfield  {author} {\bibinfo {author} {\bibfnamefont {A.}~\bibnamefont
  {Nandy}}\ and\ \bibinfo {author} {\bibfnamefont {A.}~\bibnamefont
  {Chakrabarti}},\ }\href@noop {} {\bibfield  {journal} {\bibinfo  {journal}
  {Physics Letters A}\ }\textbf {\bibinfo {volume} {379}},\ \bibinfo {pages}
  {2876} (\bibinfo {year} {2015})}\BibitemShut {NoStop}%
\bibitem [{\citenamefont {Nandy}\ \emph {et~al.}(2015)\citenamefont {Nandy},
  \citenamefont {Pal},\ and\ \citenamefont {Chakrabarti}}]{nandy2015flat}%
  \BibitemOpen
  \bibfield  {author} {\bibinfo {author} {\bibfnamefont {A.}~\bibnamefont
  {Nandy}}, \bibinfo {author} {\bibfnamefont {B.}~\bibnamefont {Pal}}, \ and\
  \bibinfo {author} {\bibfnamefont {A.}~\bibnamefont {Chakrabarti}},\
  }\href@noop {} {\bibfield  {journal} {\bibinfo  {journal} {Journal of
  Physics: Condensed Matter}\ }\textbf {\bibinfo {volume} {27}},\ \bibinfo
  {pages} {125501} (\bibinfo {year} {2015})}\BibitemShut {NoStop}%
\bibitem [{\citenamefont {Pal}\ and\ \citenamefont {Saha}(2018)}]{pal2018flat}%
  \BibitemOpen
  \bibfield  {author} {\bibinfo {author} {\bibfnamefont {B.}~\bibnamefont
  {Pal}}\ and\ \bibinfo {author} {\bibfnamefont {K.}~\bibnamefont {Saha}},\
  }\href@noop {} {\bibfield  {journal} {\bibinfo  {journal} {Physical Review
  B}\ }\textbf {\bibinfo {volume} {97}},\ \bibinfo {pages} {195101} (\bibinfo
  {year} {2018})}\BibitemShut {NoStop}%
\bibitem [{\citenamefont {Iliasov}\ \emph {et~al.}(2019)\citenamefont
  {Iliasov}, \citenamefont {Katsnelson},\ and\ \citenamefont
  {Yuan}}]{iliasov2019power}%
  \BibitemOpen
  \bibfield  {author} {\bibinfo {author} {\bibfnamefont {A.~A.}\ \bibnamefont
  {Iliasov}}, \bibinfo {author} {\bibfnamefont {M.~I.}\ \bibnamefont
  {Katsnelson}}, \ and\ \bibinfo {author} {\bibfnamefont {S.}~\bibnamefont
  {Yuan}},\ }\href@noop {} {\bibfield  {journal} {\bibinfo  {journal} {Phys.
  Rev. B}\ }\textbf {\bibinfo {volume} {99}},\ \bibinfo {pages} {075402}
  (\bibinfo {year} {2019})}\BibitemShut {NoStop}%
\bibitem [{\citenamefont {Brzezi{\'n}ska}\ \emph {et~al.}(2018)\citenamefont
  {Brzezi{\'n}ska}, \citenamefont {Cook},\ and\ \citenamefont
  {Neupert}}]{brzezinska2018topology}%
  \BibitemOpen
  \bibfield  {author} {\bibinfo {author} {\bibfnamefont {M.}~\bibnamefont
  {Brzezi{\'n}ska}}, \bibinfo {author} {\bibfnamefont {A.~M.}\ \bibnamefont
  {Cook}}, \ and\ \bibinfo {author} {\bibfnamefont {T.}~\bibnamefont
  {Neupert}},\ }\href@noop {} {\bibfield  {journal} {\bibinfo  {journal} {Phys.
  Rev. B}\ }\textbf {\bibinfo {volume} {98}},\ \bibinfo {pages} {205116}
  (\bibinfo {year} {2018})}\BibitemShut {NoStop}%
\bibitem [{\citenamefont {Pai}\ and\ \citenamefont
  {Prem}(2019)}]{pai2019topological}%
  \BibitemOpen
  \bibfield  {author} {\bibinfo {author} {\bibfnamefont {S.}~\bibnamefont
  {Pai}}\ and\ \bibinfo {author} {\bibfnamefont {A.}~\bibnamefont {Prem}},\
  }\href@noop {} {\bibfield  {journal} {\bibinfo  {journal} {Physical Review
  B}\ }\textbf {\bibinfo {volume} {100}},\ \bibinfo {pages} {155135} (\bibinfo
  {year} {2019})}\BibitemShut {NoStop}%
\bibitem [{\citenamefont {Manna}\ \emph {et~al.}(2020)\citenamefont {Manna},
  \citenamefont {Pal}, \citenamefont {Wang}, \citenamefont {Nielsen} \emph
  {et~al.}}]{manna2020anyons}%
  \BibitemOpen
  \bibfield  {author} {\bibinfo {author} {\bibfnamefont {S.}~\bibnamefont
  {Manna}}, \bibinfo {author} {\bibfnamefont {B.}~\bibnamefont {Pal}}, \bibinfo
  {author} {\bibfnamefont {W.}~\bibnamefont {Wang}}, \bibinfo {author}
  {\bibfnamefont {A.~E.~B.}\ \bibnamefont {Nielsen}},  \emph {et~al.},\
  }\href@noop {} {\bibfield  {journal} {\bibinfo  {journal} {Physical Review
  Research}\ }\textbf {\bibinfo {volume} {2}},\ \bibinfo {pages} {023401}
  (\bibinfo {year} {2020})}\BibitemShut {NoStop}%
\bibitem [{\citenamefont {Pedersen}\ \emph
  {et~al.}(2008{\natexlab{a}})\citenamefont {Pedersen}, \citenamefont {Flindt},
  \citenamefont {Pedersen}, \citenamefont {Mortensen}, \citenamefont {Jauho},\
  and\ \citenamefont {Pedersen}}]{pedersen2008graphene}%
  \BibitemOpen
  \bibfield  {author} {\bibinfo {author} {\bibfnamefont {T.~G.}\ \bibnamefont
  {Pedersen}}, \bibinfo {author} {\bibfnamefont {C.}~\bibnamefont {Flindt}},
  \bibinfo {author} {\bibfnamefont {J.}~\bibnamefont {Pedersen}}, \bibinfo
  {author} {\bibfnamefont {N.~A.}\ \bibnamefont {Mortensen}}, \bibinfo {author}
  {\bibfnamefont {A.-P.}\ \bibnamefont {Jauho}}, \ and\ \bibinfo {author}
  {\bibfnamefont {K.}~\bibnamefont {Pedersen}},\ }\href@noop {} {\bibfield
  {journal} {\bibinfo  {journal} {Physical Review Letters}\ }\textbf {\bibinfo
  {volume} {100}},\ \bibinfo {pages} {136804} (\bibinfo {year}
  {2008}{\natexlab{a}})}\BibitemShut {NoStop}%
\bibitem [{\citenamefont {Pedersen}\ \emph
  {et~al.}(2008{\natexlab{b}})\citenamefont {Pedersen}, \citenamefont {Flindt},
  \citenamefont {Pedersen}, \citenamefont {Jauho}, \citenamefont {Mortensen},\
  and\ \citenamefont {Pedersen}}]{pedersen2008optical}%
  \BibitemOpen
  \bibfield  {author} {\bibinfo {author} {\bibfnamefont {T.~G.}\ \bibnamefont
  {Pedersen}}, \bibinfo {author} {\bibfnamefont {C.}~\bibnamefont {Flindt}},
  \bibinfo {author} {\bibfnamefont {J.}~\bibnamefont {Pedersen}}, \bibinfo
  {author} {\bibfnamefont {A.-P.}\ \bibnamefont {Jauho}}, \bibinfo {author}
  {\bibfnamefont {N.~A.}\ \bibnamefont {Mortensen}}, \ and\ \bibinfo {author}
  {\bibfnamefont {K.}~\bibnamefont {Pedersen}},\ }\href@noop {} {\bibfield
  {journal} {\bibinfo  {journal} {Physical Review B}\ }\textbf {\bibinfo
  {volume} {77}},\ \bibinfo {pages} {245431} (\bibinfo {year}
  {2008}{\natexlab{b}})}\BibitemShut {NoStop}%
\bibitem [{\citenamefont {F{\"u}rst}\ \emph {et~al.}(2009)\citenamefont
  {F{\"u}rst}, \citenamefont {Pedersen}, \citenamefont {Flindt}, \citenamefont
  {Mortensen}, \citenamefont {Brandbyge}, \citenamefont {Pedersen},\ and\
  \citenamefont {Jauho}}]{furst2009electronic}%
  \BibitemOpen
  \bibfield  {author} {\bibinfo {author} {\bibfnamefont {J.~A.}\ \bibnamefont
  {F{\"u}rst}}, \bibinfo {author} {\bibfnamefont {J.~G.}\ \bibnamefont
  {Pedersen}}, \bibinfo {author} {\bibfnamefont {C.}~\bibnamefont {Flindt}},
  \bibinfo {author} {\bibfnamefont {N.~A.}\ \bibnamefont {Mortensen}}, \bibinfo
  {author} {\bibfnamefont {M.}~\bibnamefont {Brandbyge}}, \bibinfo {author}
  {\bibfnamefont {T.~G.}\ \bibnamefont {Pedersen}}, \ and\ \bibinfo {author}
  {\bibfnamefont {A.-P.}\ \bibnamefont {Jauho}},\ }\href@noop {} {\bibfield
  {journal} {\bibinfo  {journal} {New J. Phys.}\ }\textbf {\bibinfo {volume}
  {11}},\ \bibinfo {pages} {095020} (\bibinfo {year} {2009})}\BibitemShut
  {NoStop}%
\bibitem [{\citenamefont {Gunst}\ \emph {et~al.}(2011)\citenamefont {Gunst},
  \citenamefont {Markussen}, \citenamefont {Jauho},\ and\ \citenamefont
  {Brandbyge}}]{gunst2011thermoelectric}%
  \BibitemOpen
  \bibfield  {author} {\bibinfo {author} {\bibfnamefont {T.}~\bibnamefont
  {Gunst}}, \bibinfo {author} {\bibfnamefont {T.}~\bibnamefont {Markussen}},
  \bibinfo {author} {\bibfnamefont {A.-P.}\ \bibnamefont {Jauho}}, \ and\
  \bibinfo {author} {\bibfnamefont {M.}~\bibnamefont {Brandbyge}},\ }\href@noop
  {} {\bibfield  {journal} {\bibinfo  {journal} {Physical Review B}\ }\textbf
  {\bibinfo {volume} {84}},\ \bibinfo {pages} {155449} (\bibinfo {year}
  {2011})}\BibitemShut {NoStop}%
\bibitem [{\citenamefont {Yuan}\ \emph {et~al.}(2013)\citenamefont {Yuan},
  \citenamefont {Rold{\'a}n}, \citenamefont {Jauho},\ and\ \citenamefont
  {Katsnelson}}]{yuan2013electronic}%
  \BibitemOpen
  \bibfield  {author} {\bibinfo {author} {\bibfnamefont {S.}~\bibnamefont
  {Yuan}}, \bibinfo {author} {\bibfnamefont {R.}~\bibnamefont {Rold{\'a}n}},
  \bibinfo {author} {\bibfnamefont {A.-P.}\ \bibnamefont {Jauho}}, \ and\
  \bibinfo {author} {\bibfnamefont {M.~I.}\ \bibnamefont {Katsnelson}},\
  }\href@noop {} {\bibfield  {journal} {\bibinfo  {journal} {Physical Review
  B}\ }\textbf {\bibinfo {volume} {87}},\ \bibinfo {pages} {085430} (\bibinfo
  {year} {2013})}\BibitemShut {NoStop}%
\bibitem [{\citenamefont {Mandal}\ \emph {et~al.}(2013)\citenamefont {Mandal},
  \citenamefont {Laha}, \citenamefont {Das}, \citenamefont {Saha},
  \citenamefont {Barman}, \citenamefont {Raychaudhuri},\ and\ \citenamefont
  {Barman}}]{mandal2013effects}%
  \BibitemOpen
  \bibfield  {author} {\bibinfo {author} {\bibfnamefont {R.}~\bibnamefont
  {Mandal}}, \bibinfo {author} {\bibfnamefont {P.}~\bibnamefont {Laha}},
  \bibinfo {author} {\bibfnamefont {K.}~\bibnamefont {Das}}, \bibinfo {author}
  {\bibfnamefont {S.}~\bibnamefont {Saha}}, \bibinfo {author} {\bibfnamefont
  {S.}~\bibnamefont {Barman}}, \bibinfo {author} {\bibfnamefont
  {A.}~\bibnamefont {Raychaudhuri}}, \ and\ \bibinfo {author} {\bibfnamefont
  {A.}~\bibnamefont {Barman}},\ }\href@noop {} {\bibfield  {journal} {\bibinfo
  {journal} {Applied Physics Letters}\ }\textbf {\bibinfo {volume} {103}},\
  \bibinfo {pages} {262410} (\bibinfo {year} {2013})}\BibitemShut {NoStop}%
\bibitem [{\citenamefont {Cupo}\ \emph {et~al.}(2017)\citenamefont {Cupo},
  \citenamefont {Masih~Das}, \citenamefont {Chien}, \citenamefont {Danda},
  \citenamefont {Kharche}, \citenamefont {Tristant}, \citenamefont
  {Drndi\'{c}},\ and\ \citenamefont {Meunier}}]{cupo2017periodic}%
  \BibitemOpen
  \bibfield  {author} {\bibinfo {author} {\bibfnamefont {A.}~\bibnamefont
  {Cupo}}, \bibinfo {author} {\bibfnamefont {P.}~\bibnamefont {Masih~Das}},
  \bibinfo {author} {\bibfnamefont {C.-C.}\ \bibnamefont {Chien}}, \bibinfo
  {author} {\bibfnamefont {G.}~\bibnamefont {Danda}}, \bibinfo {author}
  {\bibfnamefont {N.}~\bibnamefont {Kharche}}, \bibinfo {author} {\bibfnamefont
  {D.}~\bibnamefont {Tristant}}, \bibinfo {author} {\bibfnamefont
  {M.}~\bibnamefont {Drndi\'{c}}}, \ and\ \bibinfo {author} {\bibfnamefont
  {V.}~\bibnamefont {Meunier}},\ }\href@noop {} {\bibfield  {journal} {\bibinfo
   {journal} {ACS nano}\ }\textbf {\bibinfo {volume} {11}},\ \bibinfo {pages}
  {7494} (\bibinfo {year} {2017})}\BibitemShut {NoStop}%
\bibitem [{\citenamefont {Eroms}\ and\ \citenamefont
  {Weiss}(2009)}]{eroms2009weak}%
  \BibitemOpen
  \bibfield  {author} {\bibinfo {author} {\bibfnamefont {J.}~\bibnamefont
  {Eroms}}\ and\ \bibinfo {author} {\bibfnamefont {D.}~\bibnamefont {Weiss}},\
  }\href@noop {} {\bibfield  {journal} {\bibinfo  {journal} {New Journal of
  Physics}\ }\textbf {\bibinfo {volume} {11}},\ \bibinfo {pages} {095021}
  (\bibinfo {year} {2009})}\BibitemShut {NoStop}%
\bibitem [{\citenamefont {Mackenzie}\ \emph {et~al.}(2017)\citenamefont
  {Mackenzie}, \citenamefont {Cagliani}, \citenamefont {Gammelgaard},
  \citenamefont {Jessen}, \citenamefont {Petersen},\ and\ \citenamefont
  {B{\o}ggild}}]{mackenzie2017graphene}%
  \BibitemOpen
  \bibfield  {author} {\bibinfo {author} {\bibfnamefont {D.~M.}\ \bibnamefont
  {Mackenzie}}, \bibinfo {author} {\bibfnamefont {A.}~\bibnamefont {Cagliani}},
  \bibinfo {author} {\bibfnamefont {L.}~\bibnamefont {Gammelgaard}}, \bibinfo
  {author} {\bibfnamefont {B.~S.}\ \bibnamefont {Jessen}}, \bibinfo {author}
  {\bibfnamefont {D.~H.}\ \bibnamefont {Petersen}}, \ and\ \bibinfo {author}
  {\bibfnamefont {P.}~\bibnamefont {B{\o}ggild}},\ }\href@noop {} {\bibfield
  {journal} {\bibinfo  {journal} {International Journal of Nanotechnology}\
  }\textbf {\bibinfo {volume} {14}},\ \bibinfo {pages} {226} (\bibinfo {year}
  {2017})}\BibitemShut {NoStop}%
\bibitem [{\citenamefont {Xu}\ \emph {et~al.}(2019)\citenamefont {Xu},
  \citenamefont {Tang}, \citenamefont {Du},\ and\ \citenamefont
  {Hao}}]{xu2019detecting}%
  \BibitemOpen
  \bibfield  {author} {\bibinfo {author} {\bibfnamefont {D.}~\bibnamefont
  {Xu}}, \bibinfo {author} {\bibfnamefont {S.}~\bibnamefont {Tang}}, \bibinfo
  {author} {\bibfnamefont {X.}~\bibnamefont {Du}}, \ and\ \bibinfo {author}
  {\bibfnamefont {Q.}~\bibnamefont {Hao}},\ }\href@noop {} {\bibfield
  {journal} {\bibinfo  {journal} {Carbon}\ }\textbf {\bibinfo {volume} {144}},\
  \bibinfo {pages} {601} (\bibinfo {year} {2019})}\BibitemShut {NoStop}%
\bibitem [{\citenamefont {Choudhury}\ \emph {et~al.}(2019)\citenamefont
  {Choudhury}, \citenamefont {Barman}, \citenamefont {Otani},\ and\
  \citenamefont {Barman}}]{choudhury2019controlled}%
  \BibitemOpen
  \bibfield  {author} {\bibinfo {author} {\bibfnamefont {S.}~\bibnamefont
  {Choudhury}}, \bibinfo {author} {\bibfnamefont {S.}~\bibnamefont {Barman}},
  \bibinfo {author} {\bibfnamefont {Y.}~\bibnamefont {Otani}}, \ and\ \bibinfo
  {author} {\bibfnamefont {A.}~\bibnamefont {Barman}},\ }\href@noop {}
  {\bibfield  {journal} {\bibinfo  {journal} {J. Magn. Magn. Mater.}\ }\textbf
  {\bibinfo {volume} {489}},\ \bibinfo {pages} {165408} (\bibinfo {year}
  {2019})}\BibitemShut {NoStop}%
\bibitem [{\citenamefont {Terrones}\ \emph {et~al.}(2012)\citenamefont
  {Terrones}, \citenamefont {Lv}, \citenamefont {Terrones},\ and\ \citenamefont
  {Dresselhaus}}]{terrones2012role}%
  \BibitemOpen
  \bibfield  {author} {\bibinfo {author} {\bibfnamefont {H.}~\bibnamefont
  {Terrones}}, \bibinfo {author} {\bibfnamefont {R.}~\bibnamefont {Lv}},
  \bibinfo {author} {\bibfnamefont {M.}~\bibnamefont {Terrones}}, \ and\
  \bibinfo {author} {\bibfnamefont {M.~S.}\ \bibnamefont {Dresselhaus}},\
  }\href@noop {} {\bibfield  {journal} {\bibinfo  {journal} {Rep. Prog. Phys.}\
  }\textbf {\bibinfo {volume} {75}},\ \bibinfo {pages} {062501} (\bibinfo
  {year} {2012})}\BibitemShut {NoStop}%
\bibitem [{\citenamefont {Wang}\ \emph {et~al.}(2012)\citenamefont {Wang},
  \citenamefont {Wang}, \citenamefont {Cheng}, \citenamefont {Li},
  \citenamefont {Yao}, \citenamefont {Zhang}, \citenamefont {Dong},
  \citenamefont {Wang}, \citenamefont {Schwingenschl\"{o}gl}, \citenamefont
  {Yang} \emph {et~al.}}]{wang2012doping}%
  \BibitemOpen
  \bibfield  {author} {\bibinfo {author} {\bibfnamefont {H.}~\bibnamefont
  {Wang}}, \bibinfo {author} {\bibfnamefont {Q.}~\bibnamefont {Wang}}, \bibinfo
  {author} {\bibfnamefont {Y.}~\bibnamefont {Cheng}}, \bibinfo {author}
  {\bibfnamefont {K.}~\bibnamefont {Li}}, \bibinfo {author} {\bibfnamefont
  {Y.}~\bibnamefont {Yao}}, \bibinfo {author} {\bibfnamefont {Q.}~\bibnamefont
  {Zhang}}, \bibinfo {author} {\bibfnamefont {C.}~\bibnamefont {Dong}},
  \bibinfo {author} {\bibfnamefont {P.}~\bibnamefont {Wang}}, \bibinfo {author}
  {\bibfnamefont {U.}~\bibnamefont {Schwingenschl\"{o}gl}}, \bibinfo {author}
  {\bibfnamefont {W.}~\bibnamefont {Yang}},  \emph {et~al.},\ }\href@noop {}
  {\bibfield  {journal} {\bibinfo  {journal} {Nano Lett.}\ }\textbf {\bibinfo
  {volume} {12}},\ \bibinfo {pages} {141} (\bibinfo {year} {2012})}\BibitemShut
  {NoStop}%
\bibitem [{\citenamefont {Komsa}\ \emph {et~al.}(2012)\citenamefont {Komsa},
  \citenamefont {Kotakoski}, \citenamefont {Kurasch}, \citenamefont {Lehtinen},
  \citenamefont {Kaiser},\ and\ \citenamefont {Krasheninnikov}}]{komsa2012two}%
  \BibitemOpen
  \bibfield  {author} {\bibinfo {author} {\bibfnamefont {H.-P.}\ \bibnamefont
  {Komsa}}, \bibinfo {author} {\bibfnamefont {J.}~\bibnamefont {Kotakoski}},
  \bibinfo {author} {\bibfnamefont {S.}~\bibnamefont {Kurasch}}, \bibinfo
  {author} {\bibfnamefont {O.}~\bibnamefont {Lehtinen}}, \bibinfo {author}
  {\bibfnamefont {U.}~\bibnamefont {Kaiser}}, \ and\ \bibinfo {author}
  {\bibfnamefont {A.~V.}\ \bibnamefont {Krasheninnikov}},\ }\href@noop {}
  {\bibfield  {journal} {\bibinfo  {journal} {Phys. Rev. Lett.}\ }\textbf
  {\bibinfo {volume} {109}},\ \bibinfo {pages} {035503} (\bibinfo {year}
  {2012})}\BibitemShut {NoStop}%
\bibitem [{\citenamefont {Feng}\ \emph {et~al.}(2017)\citenamefont {Feng},
  \citenamefont {Lin}, \citenamefont {Gan}, \citenamefont {Lv},\ and\
  \citenamefont {Terrones}}]{feng2017doping}%
  \BibitemOpen
  \bibfield  {author} {\bibinfo {author} {\bibfnamefont {S.}~\bibnamefont
  {Feng}}, \bibinfo {author} {\bibfnamefont {Z.}~\bibnamefont {Lin}}, \bibinfo
  {author} {\bibfnamefont {X.}~\bibnamefont {Gan}}, \bibinfo {author}
  {\bibfnamefont {R.}~\bibnamefont {Lv}}, \ and\ \bibinfo {author}
  {\bibfnamefont {M.}~\bibnamefont {Terrones}},\ }\href@noop {} {\bibfield
  {journal} {\bibinfo  {journal} {Nanoscale Horiz.}\ }\textbf {\bibinfo
  {volume} {2}},\ \bibinfo {pages} {72} (\bibinfo {year} {2017})}\BibitemShut
  {NoStop}%
\bibitem [{\citenamefont {Zhao}\ \emph {et~al.}(2013)\citenamefont {Zhao},
  \citenamefont {Levendorf}, \citenamefont {Goncher}, \citenamefont {Schiros},
  \citenamefont {Palova}, \citenamefont {Zabet-Khosousi}, \citenamefont {Rim},
  \citenamefont {Gutierrez}, \citenamefont {Nordlund}, \citenamefont {Jaye}
  \emph {et~al.}}]{zhao2013local}%
  \BibitemOpen
  \bibfield  {author} {\bibinfo {author} {\bibfnamefont {L.}~\bibnamefont
  {Zhao}}, \bibinfo {author} {\bibfnamefont {M.}~\bibnamefont {Levendorf}},
  \bibinfo {author} {\bibfnamefont {S.}~\bibnamefont {Goncher}}, \bibinfo
  {author} {\bibfnamefont {T.}~\bibnamefont {Schiros}}, \bibinfo {author}
  {\bibfnamefont {L.}~\bibnamefont {Palova}}, \bibinfo {author} {\bibfnamefont
  {A.}~\bibnamefont {Zabet-Khosousi}}, \bibinfo {author} {\bibfnamefont
  {K.~T.}\ \bibnamefont {Rim}}, \bibinfo {author} {\bibfnamefont
  {C.}~\bibnamefont {Gutierrez}}, \bibinfo {author} {\bibfnamefont
  {D.}~\bibnamefont {Nordlund}}, \bibinfo {author} {\bibfnamefont
  {C.}~\bibnamefont {Jaye}},  \emph {et~al.},\ }\href@noop {} {\bibfield
  {journal} {\bibinfo  {journal} {Nano Lett.}\ }\textbf {\bibinfo {volume}
  {13}},\ \bibinfo {pages} {4659} (\bibinfo {year} {2013})}\BibitemShut
  {NoStop}%
\bibitem [{\citenamefont {Zhao}\ and\ \citenamefont
  {Chen}(2011)}]{zhao2011electronic}%
  \BibitemOpen
  \bibfield  {author} {\bibinfo {author} {\bibfnamefont {P.-L.}\ \bibnamefont
  {Zhao}}\ and\ \bibinfo {author} {\bibfnamefont {X.}~\bibnamefont {Chen}},\
  }\href@noop {} {\bibfield  {journal} {\bibinfo  {journal} {Applied Physics
  Letters}\ }\textbf {\bibinfo {volume} {99}},\ \bibinfo {pages} {182108}
  (\bibinfo {year} {2011})}\BibitemShut {NoStop}%
\bibitem [{\citenamefont {Padilha}\ \emph {et~al.}(2015)\citenamefont
  {Padilha}, \citenamefont {Fazzio},\ and\ \citenamefont
  {da~Silva}}]{padilha2015van}%
  \BibitemOpen
  \bibfield  {author} {\bibinfo {author} {\bibfnamefont {J.~E.}\ \bibnamefont
  {Padilha}}, \bibinfo {author} {\bibfnamefont {A.}~\bibnamefont {Fazzio}}, \
  and\ \bibinfo {author} {\bibfnamefont {A.~J.~R.}\ \bibnamefont {da~Silva}},\
  }\href@noop {} {\bibfield  {journal} {\bibinfo  {journal} {Phys. Rev. Lett.}\
  }\textbf {\bibinfo {volume} {114}},\ \bibinfo {pages} {066803} (\bibinfo
  {year} {2015})}\BibitemShut {NoStop}%
\bibitem [{\citenamefont {Santos}\ and\ \citenamefont
  {Kaxiras}(2013)}]{santos2013electric}%
  \BibitemOpen
  \bibfield  {author} {\bibinfo {author} {\bibfnamefont {E.~J.}\ \bibnamefont
  {Santos}}\ and\ \bibinfo {author} {\bibfnamefont {E.}~\bibnamefont
  {Kaxiras}},\ }\href@noop {} {\bibfield  {journal} {\bibinfo  {journal} {Nano
  letters}\ }\textbf {\bibinfo {volume} {13}},\ \bibinfo {pages} {898}
  (\bibinfo {year} {2013})}\BibitemShut {NoStop}%
\bibitem [{\citenamefont {Chen}\ \emph {et~al.}(2020)\citenamefont {Chen},
  \citenamefont {Kraft}, \citenamefont {Danneau}, \citenamefont {Richter},\
  and\ \citenamefont {Liu}}]{chen2020electrostatic}%
  \BibitemOpen
  \bibfield  {author} {\bibinfo {author} {\bibfnamefont {S.-C.}\ \bibnamefont
  {Chen}}, \bibinfo {author} {\bibfnamefont {R.}~\bibnamefont {Kraft}},
  \bibinfo {author} {\bibfnamefont {R.}~\bibnamefont {Danneau}}, \bibinfo
  {author} {\bibfnamefont {K.}~\bibnamefont {Richter}}, \ and\ \bibinfo
  {author} {\bibfnamefont {M.-H.}\ \bibnamefont {Liu}},\ }\href@noop {}
  {\bibfield  {journal} {\bibinfo  {journal} {Communications Physics}\ }\textbf
  {\bibinfo {volume} {3}},\ \bibinfo {pages} {1} (\bibinfo {year}
  {2020})}\BibitemShut {NoStop}%
\bibitem [{\citenamefont {Huber}\ \emph {et~al.}(2020)\citenamefont {Huber},
  \citenamefont {Liu}, \citenamefont {Chen}, \citenamefont {Drienovsky},
  \citenamefont {Sandner}, \citenamefont {Watanabe}, \citenamefont {Taniguchi},
  \citenamefont {Richter}, \citenamefont {Weiss},\ and\ \citenamefont
  {Eroms}}]{huber2020tunable}%
  \BibitemOpen
  \bibfield  {author} {\bibinfo {author} {\bibfnamefont {R.}~\bibnamefont
  {Huber}}, \bibinfo {author} {\bibfnamefont {M.-H.}\ \bibnamefont {Liu}},
  \bibinfo {author} {\bibfnamefont {S.-C.}\ \bibnamefont {Chen}}, \bibinfo
  {author} {\bibfnamefont {M.}~\bibnamefont {Drienovsky}}, \bibinfo {author}
  {\bibfnamefont {A.}~\bibnamefont {Sandner}}, \bibinfo {author} {\bibfnamefont
  {K.}~\bibnamefont {Watanabe}}, \bibinfo {author} {\bibfnamefont
  {T.}~\bibnamefont {Taniguchi}}, \bibinfo {author} {\bibfnamefont
  {K.}~\bibnamefont {Richter}}, \bibinfo {author} {\bibfnamefont
  {D.}~\bibnamefont {Weiss}}, \ and\ \bibinfo {author} {\bibfnamefont
  {J.}~\bibnamefont {Eroms}},\ }\href@noop {} {\bibfield  {journal} {\bibinfo
  {journal} {arXiv preprint arXiv:2003.07376}\ } (\bibinfo {year}
  {2020})}\BibitemShut {NoStop}%
\bibitem [{\citenamefont {Yuan}\ \emph {et~al.}(2010)\citenamefont {Yuan},
  \citenamefont {De~Raedt},\ and\ \citenamefont {Katsnelson}}]{TBPM}%
  \BibitemOpen
  \bibfield  {author} {\bibinfo {author} {\bibfnamefont {S.}~\bibnamefont
  {Yuan}}, \bibinfo {author} {\bibfnamefont {H.}~\bibnamefont {De~Raedt}}, \
  and\ \bibinfo {author} {\bibfnamefont {M.~I.}\ \bibnamefont {Katsnelson}},\
  }\href@noop {} {\bibfield  {journal} {\bibinfo  {journal} {Phys. Rev. B}\
  }\textbf {\bibinfo {volume} {82}},\ \bibinfo {pages} {115448} (\bibinfo
  {year} {2010})}\BibitemShut {NoStop}%
\bibitem [{\citenamefont {Hams}\ and\ \citenamefont
  {De~Raedt}(2000)}]{hams2000fast}%
  \BibitemOpen
  \bibfield  {author} {\bibinfo {author} {\bibfnamefont {A.}~\bibnamefont
  {Hams}}\ and\ \bibinfo {author} {\bibfnamefont {H.}~\bibnamefont
  {De~Raedt}},\ }\href@noop {} {\bibfield  {journal} {\bibinfo  {journal}
  {Physical Review E}\ }\textbf {\bibinfo {volume} {62}},\ \bibinfo {pages}
  {4365} (\bibinfo {year} {2000})}\BibitemShut {NoStop}%
\bibitem [{\citenamefont {Kosloff}\ and\ \citenamefont
  {Kosloff}(1983)}]{kosloff1983fourier}%
  \BibitemOpen
  \bibfield  {author} {\bibinfo {author} {\bibfnamefont {D.}~\bibnamefont
  {Kosloff}}\ and\ \bibinfo {author} {\bibfnamefont {R.}~\bibnamefont
  {Kosloff}},\ }\href@noop {} {\bibfield  {journal} {\bibinfo  {journal} {J.
  Comput. Phys.}\ }\textbf {\bibinfo {volume} {52}},\ \bibinfo {pages} {35}
  (\bibinfo {year} {1983})}\BibitemShut {NoStop}%
\bibitem [{\citenamefont {Groth}\ \emph {et~al.}(2014)\citenamefont {Groth},
  \citenamefont {Wimmer}, \citenamefont {Akhmerov},\ and\ \citenamefont
  {Waintal}}]{groth2014kwant}%
  \BibitemOpen
  \bibfield  {author} {\bibinfo {author} {\bibfnamefont {C.~W.}\ \bibnamefont
  {Groth}}, \bibinfo {author} {\bibfnamefont {M.}~\bibnamefont {Wimmer}},
  \bibinfo {author} {\bibfnamefont {A.~R.}\ \bibnamefont {Akhmerov}}, \ and\
  \bibinfo {author} {\bibfnamefont {X.}~\bibnamefont {Waintal}},\ }\href@noop
  {} {\bibfield  {journal} {\bibinfo  {journal} {New J. Phys.}\ }\textbf
  {\bibinfo {volume} {16}},\ \bibinfo {pages} {063065} (\bibinfo {year}
  {2014})}\BibitemShut {NoStop}%
\bibitem [{\citenamefont {Ketzmerick}(1996)}]{ketzmerick1996fractal}%
  \BibitemOpen
  \bibfield  {author} {\bibinfo {author} {\bibfnamefont {R.}~\bibnamefont
  {Ketzmerick}},\ }\href@noop {} {\bibfield  {journal} {\bibinfo  {journal}
  {Phys. Rev. B}\ }\textbf {\bibinfo {volume} {54}},\ \bibinfo {pages} {10841}
  (\bibinfo {year} {1996})}\BibitemShut {NoStop}%
\bibitem [{\citenamefont {Sachrajda}\ \emph {et~al.}(1998)\citenamefont
  {Sachrajda}, \citenamefont {Ketzmerick}, \citenamefont {Gould}, \citenamefont
  {Feng}, \citenamefont {Kelly}, \citenamefont {Delage},\ and\ \citenamefont
  {Wasilewski}}]{sachrajda1998fractal}%
  \BibitemOpen
  \bibfield  {author} {\bibinfo {author} {\bibfnamefont {A.~S.}\ \bibnamefont
  {Sachrajda}}, \bibinfo {author} {\bibfnamefont {R.}~\bibnamefont
  {Ketzmerick}}, \bibinfo {author} {\bibfnamefont {C.}~\bibnamefont {Gould}},
  \bibinfo {author} {\bibfnamefont {Y.}~\bibnamefont {Feng}}, \bibinfo {author}
  {\bibfnamefont {P.~J.}\ \bibnamefont {Kelly}}, \bibinfo {author}
  {\bibfnamefont {A.}~\bibnamefont {Delage}}, \ and\ \bibinfo {author}
  {\bibfnamefont {Z.}~\bibnamefont {Wasilewski}},\ }\href@noop {} {\bibfield
  {journal} {\bibinfo  {journal} {Phys. Rev. Lett.}\ }\textbf {\bibinfo
  {volume} {80}},\ \bibinfo {pages} {1948} (\bibinfo {year}
  {1998})}\BibitemShut {NoStop}%
\bibitem [{\citenamefont {Kotim{\"a}ki}\ \emph {et~al.}(2013)\citenamefont
  {Kotim{\"a}ki}, \citenamefont {R{\"a}s{\"a}nen}, \citenamefont {Hennig},\
  and\ \citenamefont {Heller}}]{kotimaki2013fractal}%
  \BibitemOpen
  \bibfield  {author} {\bibinfo {author} {\bibfnamefont {V.}~\bibnamefont
  {Kotim{\"a}ki}}, \bibinfo {author} {\bibfnamefont {E.}~\bibnamefont
  {R{\"a}s{\"a}nen}}, \bibinfo {author} {\bibfnamefont {H.}~\bibnamefont
  {Hennig}}, \ and\ \bibinfo {author} {\bibfnamefont {E.~J.}\ \bibnamefont
  {Heller}},\ }\href@noop {} {\bibfield  {journal} {\bibinfo  {journal} {Phys.
  Rev. E}\ }\textbf {\bibinfo {volume} {88}},\ \bibinfo {pages} {022913}
  (\bibinfo {year} {2013})}\BibitemShut {NoStop}%
\bibitem [{\citenamefont {Crook}\ \emph {et~al.}(2003)\citenamefont {Crook},
  \citenamefont {Smith}, \citenamefont {Graham}, \citenamefont {Farrer},
  \citenamefont {Beere},\ and\ \citenamefont {Ritchie}}]{crook2003imaging}%
  \BibitemOpen
  \bibfield  {author} {\bibinfo {author} {\bibfnamefont {R.}~\bibnamefont
  {Crook}}, \bibinfo {author} {\bibfnamefont {C.~G.}\ \bibnamefont {Smith}},
  \bibinfo {author} {\bibfnamefont {A.~C.}\ \bibnamefont {Graham}}, \bibinfo
  {author} {\bibfnamefont {I.}~\bibnamefont {Farrer}}, \bibinfo {author}
  {\bibfnamefont {H.~E.}\ \bibnamefont {Beere}}, \ and\ \bibinfo {author}
  {\bibfnamefont {D.~A.}\ \bibnamefont {Ritchie}},\ }\href@noop {} {\bibfield
  {journal} {\bibinfo  {journal} {Phys. Rev. Lett.}\ }\textbf {\bibinfo
  {volume} {91}},\ \bibinfo {pages} {246803} (\bibinfo {year}
  {2003})}\BibitemShut {NoStop}%
\bibitem [{\citenamefont {Hegger}\ \emph {et~al.}(1996)\citenamefont {Hegger},
  \citenamefont {Huckestein}, \citenamefont {Hecker}, \citenamefont {Janssen},
  \citenamefont {Freimuth}, \citenamefont {Reckziegel},\ and\ \citenamefont
  {Tuzinski}}]{hegger1996fractal}%
  \BibitemOpen
  \bibfield  {author} {\bibinfo {author} {\bibfnamefont {H.}~\bibnamefont
  {Hegger}}, \bibinfo {author} {\bibfnamefont {B.}~\bibnamefont {Huckestein}},
  \bibinfo {author} {\bibfnamefont {K.}~\bibnamefont {Hecker}}, \bibinfo
  {author} {\bibfnamefont {M.}~\bibnamefont {Janssen}}, \bibinfo {author}
  {\bibfnamefont {A.}~\bibnamefont {Freimuth}}, \bibinfo {author}
  {\bibfnamefont {G.}~\bibnamefont {Reckziegel}}, \ and\ \bibinfo {author}
  {\bibfnamefont {R.}~\bibnamefont {Tuzinski}},\ }\href@noop {} {\bibfield
  {journal} {\bibinfo  {journal} {Phys. Rev. Lett.}\ }\textbf {\bibinfo
  {volume} {77}},\ \bibinfo {pages} {3885} (\bibinfo {year}
  {1996})}\BibitemShut {NoStop}%
\bibitem [{\citenamefont {Marlow}\ \emph {et~al.}(2006)\citenamefont {Marlow},
  \citenamefont {Taylor}, \citenamefont {Martin}, \citenamefont {Scannell},
  \citenamefont {Linke}, \citenamefont {Fairbanks}, \citenamefont {Hall},
  \citenamefont {Shorubalko}, \citenamefont {Samuelson}, \citenamefont
  {Fromhold} \emph {et~al.}}]{marlow2006unified}%
  \BibitemOpen
  \bibfield  {author} {\bibinfo {author} {\bibfnamefont {C.~A.}\ \bibnamefont
  {Marlow}}, \bibinfo {author} {\bibfnamefont {R.~P.}\ \bibnamefont {Taylor}},
  \bibinfo {author} {\bibfnamefont {T.~P.}\ \bibnamefont {Martin}}, \bibinfo
  {author} {\bibfnamefont {B.~C.}\ \bibnamefont {Scannell}}, \bibinfo {author}
  {\bibfnamefont {H.}~\bibnamefont {Linke}}, \bibinfo {author} {\bibfnamefont
  {M.~S.}\ \bibnamefont {Fairbanks}}, \bibinfo {author} {\bibfnamefont
  {G.~D.~R.}\ \bibnamefont {Hall}}, \bibinfo {author} {\bibfnamefont
  {I.}~\bibnamefont {Shorubalko}}, \bibinfo {author} {\bibfnamefont
  {L.}~\bibnamefont {Samuelson}}, \bibinfo {author} {\bibfnamefont {T.~M.}\
  \bibnamefont {Fromhold}},  \emph {et~al.},\ }\href@noop {} {\bibfield
  {journal} {\bibinfo  {journal} {Phys. Rev. B}\ }\textbf {\bibinfo {volume}
  {73}},\ \bibinfo {pages} {195318} (\bibinfo {year} {2006})}\BibitemShut
  {NoStop}%
\bibitem [{\citenamefont {Guarneri}\ and\ \citenamefont
  {Terraneo}(2001)}]{guarneri2001fractal}%
  \BibitemOpen
  \bibfield  {author} {\bibinfo {author} {\bibfnamefont {I.}~\bibnamefont
  {Guarneri}}\ and\ \bibinfo {author} {\bibfnamefont {M.}~\bibnamefont
  {Terraneo}},\ }\href@noop {} {\bibfield  {journal} {\bibinfo  {journal}
  {Phys. Rev. E}\ }\textbf {\bibinfo {volume} {65}},\ \bibinfo {pages} {015203}
  (\bibinfo {year} {2001})}\BibitemShut {NoStop}%
\bibitem [{\citenamefont {Gefen}\ \emph {et~al.}(1984)\citenamefont {Gefen},
  \citenamefont {Aharony},\ and\ \citenamefont {Mandelbrot}}]{gefen1984phase}%
  \BibitemOpen
  \bibfield  {author} {\bibinfo {author} {\bibfnamefont {Y.}~\bibnamefont
  {Gefen}}, \bibinfo {author} {\bibfnamefont {A.}~\bibnamefont {Aharony}}, \
  and\ \bibinfo {author} {\bibfnamefont {B.~B.}\ \bibnamefont {Mandelbrot}},\
  }\href@noop {} {\bibfield  {journal} {\bibinfo  {journal} {Journal of Physics
  A: Mathematical and General}\ }\textbf {\bibinfo {volume} {17}},\ \bibinfo
  {pages} {1277} (\bibinfo {year} {1984})}\BibitemShut {NoStop}%
\end{thebibliography}%

\end{document}